%% file: main.tex
\documentclass[
  journal=pasa,
  manuscript=research-paper,
  year=2023,
]{cup-journal}

\usepackage{microtype,siunitx,booktabs}
\usepackage{makecell}
\usepackage{hyperref} 
\hypersetup{colorlinks,citecolor=blue,linkcolor=blue,urlcolor=blue}
\usepackage[nolist]{acronym}
\usepackage[version=3]{mhchem}
\usepackage{mathtools}
\usepackage{adjustbox}
\usepackage[normalem]{ulem}
\usepackage{xcolor}
\newcommand\redsout{\bgroup\markoverwith{\textcolor{red}{\rule[0.5ex]{2pt}{0.4pt}}}\ULon}
\newcommand{\chandra}{{\it Chandra}}
\newcommand{\D}{$^{\circ} $}
\newcommand{\arcmin}{$^{\prime}$}
\newcommand{\arcsec}{$^{\prime\prime}$}

\newcommand{\mujybm}{\,$\mu$Jy\,beam$^{-1}$}
\newcommand{\HI}{{H{\sc I}}}
\newcommand{\HII}{{H{\sc II}}}
\newcommand{\psr}{PSR~J1638$-$4713}
\newcommand{\dmunits}{pc\,cm$^{-3}$}

\begin{acronym}
\acro{ACIS-I}{Advanced CCD Imaging Spectrometer using the wide field chip}
\acro{ASKAP}{Australian Square Kilometre Array Pathfinder}
\acro{ARC}{Australian Research Council}
\acro{ATNF}{Australia Telescope National Facility}
\acro{CALDB}{Chandra Calibration Database}
\acro{CARTA}{Cube Analysis and Rendering Tool for Astronomy}
\acro{CASDA}{CSIRO \ac{ASKAP} Science Data Archive}
\acro{CIAO}{Chandra Interactive Analysis of Observations software}
\acro{CDA}{Chandra Data Archive}
\acro{DSPSR}{Digital Signal Processing Software for Pulsar Astronomy}
\acro{DM}{dispersion measure}
\acro{EMU}{Evolutionary Map of the Universe}
\acro{FoV}{field of view}
\acro{H.E.S.S.}{High Energy Spectroscopic System}
\acro{IR}{infrared}
\acro{ISM}{interstellar medium}
\acro{MGPS}{MeerKAT Galactic Plane Survey}
\acro{MGPS-2}{Molonglo Galactic Plane Survey epoch-2}
\acro{NASA}{National Aeronautics and Space Administration}
\acro{NS}{neutron star}
\acro{ObsID}{observation identification}
\acro{PA}{position angle}
\acro{PAF}{phased array feed}
\acro{PRESTO}{Pulsar Exploration and Search Toolkit}
\acro{PWN}{pulsar wind nebula}
\acro{RFI}{radio-frequency interference}
\acro{rms}{root-mean-square}
\acro{SB}{Scheduling block}
\acro{SN}{supernova}
\acro{SNR}{supernova remnant}
\acro{UWL}{Ultra-Wideband Low}
\acro{WISE}{Wide-Field Infrared Survey Explorer}
\end{acronym}

\title{Fast as Potoroo: Radio Continuum Detection of a Bow-Shock Pulsar Wind Nebula Powered by Pulsar J1638$-$4713}

\author{Sanja Lazarevi\'c}
\affiliation{Western Sydney University, Locked Bag 1797, Penrith South DC, NSW 2751, Australia}
\alsoaffiliation{CSIRO Space and Astronomy, Australia Telescope National Facility, PO Box 76, Epping, NSW 1710, Australia}
\alsoaffiliation{Astronomical Observatory, Volgina 7, 11060 Belgrade, Serbia}
\email[S. Lazarevi\'c]{s.lazarevic@westernsydney.edu.au}

\author{Miroslav D. Filipovi\'c}
\affiliation{Western Sydney University, Locked Bag 1797, Penrith South DC, NSW 2751, Australia}

\author{Shi Dai}
\affiliation{Western Sydney University, Locked Bag 1797, Penrith South DC, NSW 2751, Australia}

\author{Roland Kothes}
\affiliation{Dominion Radio Astrophysical Observatory, Herzberg Astronomy and Astrophysics, National Research Council Canada, PO Box 248, Penticton, BC V2A 6J9, Canada}

\author{Adeel Ahmad}
\affiliation{Western Sydney University, Locked Bag 1797, Penrith South DC, NSW 2751, Australia}

\author{Rami Z. E. Alsaberi}
\affiliation{Western Sydney University, Locked Bag 1797, Penrith South DC, NSW 2751, Australia}

\author{Joel C. F. Balzan}
\affiliation{Western Sydney University, Locked Bag 1797, Penrith South DC, NSW 2751, Australia}

\author{Luke A. Barnes}
\affiliation{Western Sydney University, Locked Bag 1797, Penrith South DC, NSW 2751, Australia}

\author{William D. Cotton}
\affiliation{National Radio Astronomy Observatory, 520 Edgemont Road, Charlottesville, VA 22903, USA}
\alsoaffiliation{South African Radio Astronomy Observatory, 2 Fir Street, Black River Park, Observatory, Cape Town 7925, South Africa}

\author{Philip G. Edwards}
\affiliation{CSIRO Space and Astronomy, Australia Telescope National Facility, PO Box 76, Epping, NSW 1710, Australia}

\author{Yjan A. Gordon}
\affiliation{Department of Physics, University of Wisconsin-Madison, 1150 University Ave, Madison, WI 53706, USA}

\author{Frank Haberl}
\affiliation{Max-Planck-Institut f\"{u}r extraterrestrische Physik, Gie{\ss}enbachstra{\ss}e 1, D-85748 Garching, Germany}

\author{Andrew M. Hopkins}
\affiliation{Australian Astronomical Optics, Macquarie University, 105 Delhi Rd, North Ryde, NSW 2113, Australia}

\author{Bärbel S. Koribalski}
\affiliation{CSIRO Space and Astronomy, Australia Telescope National Facility, PO Box 76, Epping, NSW 1710, Australia}
\alsoaffiliation{Western Sydney University, Locked Bag 1797, Penrith South DC, NSW 2751, Australia}

\author{Denis Leahy}
\affiliation{Department of Physics and Astronomy, University of Calgary, Calgary, Alberta, T2N 1N4, Canada}

\author{Chandreyee Maitra}
\affiliation{Max-Planck-Institut f\"{u}r extraterrestrische Physik, Gie{\ss}enbachstra{\ss}e 1, D-85748 Garching, Germany}

\author{Marko Mi\'ci\'c}
\affiliation{Department of Physics and Astronomy, University of Alabama, Tuscaloosa, AL 35401, USA}

\author{Gavin Rowell}
\affiliation{School of Physical Sciences, The University of Adelaide, Adelaide 5005, Australia} 

\author{Manami Sasaki}
\affiliation{Dr Karl Remeis Observatory, Erlangen Centre for Astroparticle Physics, Friedrich-Alexander-Universit\"{a}t Erlangen-N\"{u}rnberg, Sternwartstra{\ss}e 7, 96049 Bamberg, Germany}

\author{Nicholas F. H. Tothill}
\affiliation{Western Sydney University, Locked Bag 1797, Penrith South DC, NSW 2751, Australia}

\author{Grazia Umana}
\affiliation{INAF\,$-$\,Osservatorio Astrofisico di Catania, Via S. Sofia 78, I-95123, Catania, Italy} 

\author{Velibor Velovi\'c}
\affiliation{Western Sydney University, Locked Bag 1797, Penrith South DC, NSW 2751, Australia}


\begin{document}

\input{Potoroo}

\bibliography{references}


\end{document}

%% file: Potoroo.tex



\begin{abstract}

We report the discovery of a bow-shock pulsar wind nebula (PWN), named Potoroo, and the detection of a young pulsar J1638$-$4713 that powers the nebula. We present a radio continuum study of the PWN based on 20-cm observations obtained from the Australian Square Kilometre Array Pathfinder (ASKAP) and MeerKAT. \psr\ was identified using Parkes radio telescope observations at frequencies above 3\,GHz. The pulsar has the second-highest dispersion measure of all known radio pulsars (1553\,\dmunits), a spin period of 65.74\,ms and a spin-down luminosity of $\dot{E}=6.1\times10^{36}$\,erg\,s$^{-1}$. The PWN has a cometary morphology and one of the greatest projected lengths among all the observed pulsar radio tails, measuring over 21\,pc for an assumed distance of 10 kpc. The remarkably long tail and atypically steep radio spectral index are attributed to the interplay of a supernova reverse shock and the PWN. The originating supernova remnant is not known so far. We estimated the pulsar kick velocity to be in the range of $1000-2000\,km\,s^{-1}$ for ages between 23 and 10\,kyr. The X-ray counterpart found in \chandra\ data, CXOU~J163802.6$-$471358, shows the same tail morphology as the radio source but is shorter by a factor of 10. The peak of the X-ray emission is offset from the peak of the radio total intensity (Stokes\,$\rm I$) emission by approximately 4.7\arcsec, but coincides well with circularly polarised (Stokes\,$\rm V$) emission. No infrared counterpart was found. 
\end{abstract}

\section{Introduction}
\label{sec:introduction}

The death of a massive star via a core-collapse \ac{SN} explosion is a dramatic event that leaves behind a \ac{SNR} and, in some instances, a rapidly rotating \ac{NS} known as a pulsar. A large fraction of the pulsar rotational energy is carried away by a constant wind of ultra-relativistic particles. When the pulsar wind interacts with the ambient medium it abruptly slows down at the termination shock, beyond which the shocked wind material inflates a bubble referred to as a \ac{PWN} \citep[for reviews, see][]{2006ARA&A..44...17G,2017SSRv..207..175R}. Within the \ac{PWN}, the ultra-relativistic particles travel through the magnetic field and generate non-thermal emission that can be observed across a wide range of frequencies, from radio to $\gamma$-rays \citep{book2}. 

\ac{PWN}e produce strong radio emission through the synchrotron process. This radiation has a flat spectrum and is often highly linearly polarised due to the ordered magnetic field configuration of the \ac{PWN} \citep[see][]{2017ASSL..446....1K}. The synchrotron spectrum can extend to the X-ray band where it typically becomes steeper than in the radio because of the synchrotron cooling effect \citep[see][]{Kargaltsev2008}. The X-ray luminosity is generated by young and high-energy electrons, mainly depending on the current energy input from the pulsar. The radio luminosity, however, is created from the older electrons, reflecting the integrated history of \ac{PWN}e. Gamma-ray emission in the GeV and TeV ranges has also been detected from \ac{PWN}e  \citep{2018A&A...612A...2H}, attributed to inverse-Compton scattering. In addition to non-thermal radiation, \ac{PWN}e produce optical H$\alpha$ emission when they ionise the neutral ambient medium \citep{2001A&A...375.1032B,2002A&A...387.1066B}. 

Studying \ac{PWN}e provides valuable insights into pulsars, the power sources of nebulae, radiative efficiency, properties of the surrounding medium, and the physics of the wind-medium interaction. It also contributes to understanding the distribution of the natal kick velocities that \ac{NS}s acquire during the \ac{SN} implosions. 

The general appearance of the \ac{PWN}e depends on a balance between a pulsar spin-down energy loss rate and the pressure of the ambient medium. For the pulsars that are propelled through the ambient medium at supersonic velocities, the resulting ram pressure transforms the \ac{PWN} into a bow-shock. This process confines the pulsar wind in the opposite direction to that of the pulsar motion, forming a cometary-like shaped tail \citep{Kargaltsev2017}.


The evolution of \ac{PWN}e can be divided into several phases which determine the overall observational properties of these objects \citep[][]{2001ApJ...563..806B,2004A&A...420..937V}. Initially, a pulsar is located near the centre of the freely expanding \ac{SNR} and is typically born with a kick velocity significantly higher than the sound speed. Thus, the \ac{PWN} expands and the pulsar drives a shock supersonically into the cool ejecta. In the next phase, the \ac{SNR} shockwave sends a reverse shock back into the interior \citep{1999ApJS..120..299T}. The interaction between the inward moving shock and the \ac{PWN} is complex and causes the \ac{PWN} to oscillate and reverberate, possibly leading the pulsar to leave its own \ac{PWN}. The \ac{SNR} reverse shock compresses the \ac{PWN} by a large factor accompanied by a sudden increase in the magnetic field, which serves to burn off the highest energy electrons \citep{1984ApJ...283..710K,2003A&A...405..617B,2009ApJ...703..662R}. The pulsar continues to travel toward the edge of the \ac{SNR}, through hot ejecta now, and will eventually break out of its parent \ac{SNR} bubble, driving a shock into the \ac{ISM}. 

Before the launch of the \chandra\ X-ray Observatory \citep{chandra} only a handful of \ac{PWN}e had been detected, mainly because of the low angular resolution of available instruments. With improved capabilities, around 30 pulsars showing indications of supersonic motion have been identified. With the exception of the Small Magellanic Cloud DEM\,S5 \ac{PWN} \citep{2019MNRAS.486.2507A}, all of \ac{PWN}e with a bow-shock reside in our galaxy and exhibit a large variety of morphologies \citep[e.g.,][]{2014IJMPS..2860172P,2012ApJ...746..105N,2020MNRAS.496..723K}. 

In this paper, we report our radio continuum discovery of a bow-shock \ac{PWN} (hereafter referred to as Potoroo\footnote{Potoroo is a small marsupial native to Australia. The species is considered a living fossil due to its minimal changes over the past 10 million years.}) and the detection of a pulsar J1638$-$4713\ that powers the \ac{PWN}. The source field has been covered during the \chandra\ survey of the Norma Galactic spiral arm region \citep{2014ApJ...796..105F}, and we found an X-ray counterpart with galactic coordinates $l=337^{\circ}.4883$ and $b=-0^{\circ}.1453$, catalogued as CXOU~J163802.6$-$471358 \citep{2014ApJ...787..129J}.

The X-ray emission of CXOU~J163802.6$-$471358 is characterised by a cometary shape, with a point source and an elongated tail extending approximately 40\arcsec\ \citep{2014ApJ...787..129J}. A diffuse, jet-like feature was detected perpendicular to the cometary tail that spans 19.5\arcsec\ (see their Figure\,1), but its significance is close to the detection limit.


To estimate the distance to CXOU~J163802.6$-$471358, \cite{2014ApJ...787..129J} measured the hydrogen column density (N$_{H}=1.5\times$10$^{23}$cm$^{-2}$) using joint \chandra\ and \textit{XMM-Newton} \citep{2001A&A...365L...1J} observations. Comparison to another source in the same area, PWN\,HESS\,J1640$-$465 \citep{2009ApJ...706.1269L}, with a slightly lower column density and distance of 10\,kpc, led the authors to adopt the same distance to the Potoroo X-ray counterpart as a lower limit. They also reported the photon index of the X-ray source as $\Gamma$=1.1$^{+0.7}_{-0.6}$.


\cite{2014ApJ...787..129J} found no flux variation between the \chandra\ and \textit{XMM-Newton} observations, conducted five years apart. They calculated the luminosity of the entire source to be approximately $4.8\times10^{33}d_{10}^{2}$\,erg\,s$^{-1}$, with the source distance $d_{10}=d/10$\,kpc. Although a pulsar search was attempted, the high background noise and poor statistics prevented the pulsation detection. 

The \ac{MGPS-2} radio observations \citep{2007MNRAS.382..382M} revealed the counterpart to the X-ray source, showing an extended tail that is aligned with the X-ray emission. \cite{2014ApJ...787..129J} reported an offset between the X-ray and radio peaks of about 40\arcsec, and no infrared counterpart.

Although \cite{2014ApJ...787..129J} discussed other potential source types, they suggested that the detected X-ray point source was most likely a previously unknown pulsar and the extended emission was a bow-shock tail created by ram pressure. 

Our study takes advantage of large-scale radio continuum surveys obtained by new-generation telescopes such as \ac{ASKAP} and MeerKAT. These advanced instruments significantly enhance our ability to detect and investigate low surface brightness objects with detail greater than ever before. We also used the new Parkes \ac{UWL} frequency receiver system \citep{hobbs20} to search for radio pulsations. The wide frequency coverage, especially up to 4\,GHz, enables us to avoid scattering and smearing due to Potoroo's large distance.

We organised the paper as follows: Section\,\ref{sec:observations} details the instruments and data used in our study; Section\,\ref{sec:results} presents our observational results which are divided into five subsections: Potoroo morphology (Subsection\,\ref{subsec:morphology}), radio spectrum (Subsection\,\ref{subsec:spectral_analysis}), polarisation analysis (Subsection\,\ref{subsec:polarisation}), pulsar detection along with its properties (Subsection\,\ref{subsec:pulsar}) and the origin of Potoroo (Subsection\,\ref{subsec:snr}); in Section\,\ref{sec:discussion}, we discuss the results; and finally, in Section\,\ref{sec:conclusion}, we present our concluding remarks. Throughout the paper, we use J2000 coordinates.

\section{Observations and Data Processing}
\label{sec:observations}

\begin{figure*}[hbt!]
\centering
\includegraphics[width=1\textwidth]{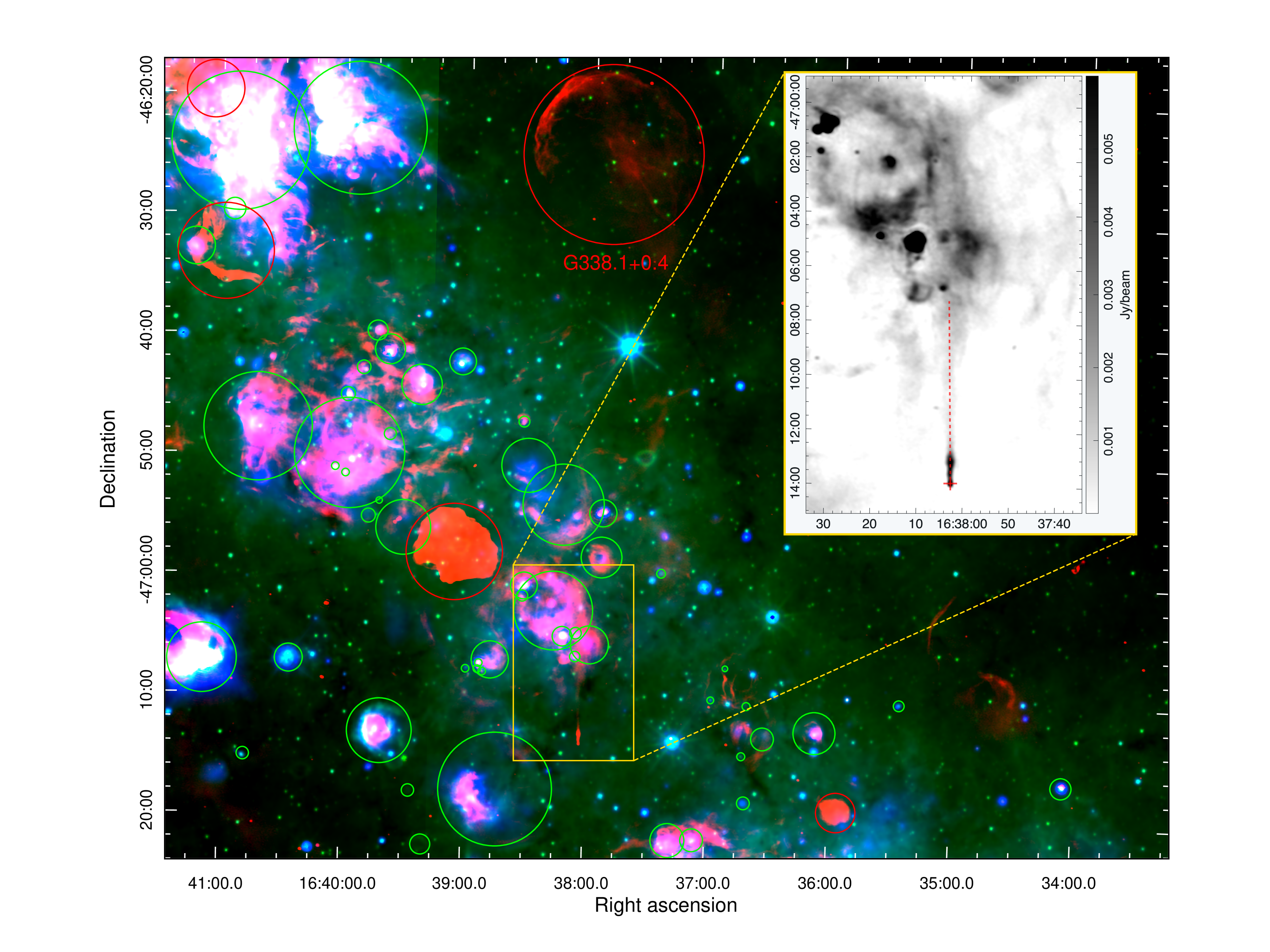}
\caption{Composite image of the Galactic plane region and Potoroo, with the red layer showing the \ac{ASKAP} total intensity image at 1368\,MHz, and the green and blue layers representing \ac{WISE} infrared images at 12\,$\mu$m and 22\,$\mu$m respectively. Known Galactic \acp{SNR} are indicated by red circles \citep{2019JApA...40...36G,Green}, while known Galactic \HII~regions are marked by green circles \citep{2014ApJS..212....1A}. The box highlights the section of deep interest. The inset is the ASKAP zoomed-in image showing Potoroo where a red cross marks the position of the X-ray source, while a red dashed line is Potoroo's axis of symmetry, which corresponds to the tail length studied in this paper.}
\label{fig:rgb}
\end{figure*}

Potoroo was detected in radio-continuum surveys obtained from \ac{ASKAP} (Figure\,\ref{fig:rgb}) and MeerKAT, as well as in the X-ray data from the \chandra\ Observatory. High time-resolution, pulsar search-mode observations were carried out using the Parkes \ac{UWL} receiver system. A summary of observations is given in Table\,\ref{tab:observation_summary}. The source field has also been covered by the infrared survey from the \ac{WISE}. We make use of this survey to better understand the surrounding \ac{ISM} and probe the origin of Potoroo. 

\begin{table}
\caption{Observational details of the Potoroo data used in this work.}
\label{tab:observation_summary}
\begin{tabular}{c c c c c} 
\toprule  
Telescope &Date&\makecell{Frequency\\ \lbrack MHz\rbrack}&\makecell{Band\\ \lbrack MHz\rbrack}&\makecell{Resolution\\ \lbrack\arcsec$\times$\arcsec\rbrack}\\ 
\hline
\hline
\ac{ASKAP} & 2021\,Sep\,11 & 944 & 800\,$-$\,1088  & 16.3\,$\times$\,13.3\\ 
\ac{ASKAP} & 2022\,Mar\,5~~ & 1368 & 1224\,$-$\,1512  & 8.8\,$\times$\,7.4\\
MeerKAT & 2018\,Aug\,26 & 1284 & 900\,$-$\,1670 &  8\,$\times$\,8 \\
Parkes  & 2019 Sep\,10 & $\approx$3000 & 704\,$-$\,4032 & 468\,$\times$\,468\\
Parkes  & 2022\,Jul\,12 & $\approx$3000 & 704\,$-$\,4032 & 468\,$\times$\,468\\
Parkes  & 2022\,Oct\,12 & $\approx$3000 & 704\,$-$\,4032 & 468\,$\times$\,468\\
\chandra\ & 2011\,Jun\,13 & $-$ & 2\,$-$\,8\footnote[1]{\chandra\ energy band is in units of keV. \label{chandra_note}} & 2.4\,$\times$\,2.4\\
\end{tabular}
\end{table}


\subsection{The Australian Square Kilometre Array Pathfinder}
\label{subsec:askap}

\ac{ASKAP} \citep{2008ExA....22..151J, 2016PASA...33...42M,2021PASA...38....9H} observed Potoroo at two central frequencies: 944\,MHz and 1368\,MHz, both with the full instantaneous bandwidth of 288\,MHz in continuum mode divided into 1\,MHz wide frequency channels (288 channels across the whole band). \ac{SB} identifications are 32043 and 37909, respectively. \ac{SB}\,32043 was obtained on 11\,September\,2021, as part of a pilot program for the \ac{EMU} survey \citep{emu,norris20}, using 34 out of 36 available \ac{ASKAP} antennas. The second block \ac{SB}\,37909 was obtained on 5\,March 2022, as a technical test designed to demonstrate the capability of the telescope \citep{test}. During that observation, 35 antennas were operating. \ac{SB}\,37909 has full polarisation data, while for \ac{SB}\,32043, only Stokes\,$\rm I$ and $\rm V$ maps are available.

Each \ac{ASKAP} antenna is 12\,m in diameter and is equipped with a \ac{PAF} beam \citep{2012SPIE.8444E..2AS} mounted at the primary focus, yielding about 30\,deg$^2$ \ac{FoV}. The 36 beams formed from the \ac{PAF} elements had a hexagonal arrangement, known as ``closepack36'', with a footprint rotation of 45\D. The pitch angle, the spacing between beams, was set to 0.90\D\ providing an approximately uniform sensitivity over the \ac{FoV} without interleaving \citep{norris20}. 

The final 944\,MHz mosaic image has a synthesised beam of 16.3\arcsec$\times$13.3\arcsec\ at a \ac{PA} of 84.5\D\ with a local \ac{rms} noise level of 131\mujybm. The 1368\,MHz mosaic image has a beam of 8.8\arcsec$\times$7.4\arcsec\ at a \ac{PA} of 84.5\D\ with \ac{rms} noise 68\mujybm. Both observations were performed with 10\,hours of exposure time. Data calibration and imaging were carried out using the ASKAPsoft data processing pipeline \citep{Guzman_Askapsoft}, running at the Pawsey Supercomputing Research Centre.

\subsection{MeerKAT}
\label{subsec:meerkat}

The MeerKAT array telescope is part of South Africa's radio observatory \citep{2009IEEEP..97.1522J,meerKAT}. The array consists of 64 antennas, each having a 13.5\,m diameter. 

We used data taken on 26\,August 2018 at a central frequency of 1284\,MHz and bandwidth of 770\,MHz (L-band receiver). The observation is part of the \ac{MGPS} that surveys the entire southern Galactic Plane within the latitude range $\pm$1.5\D (Goedhart et al., 2023, in prep.) Observations were made in a hexagonal pattern with an offset between centres of 29.6\arcmin\ for relatively uniform sensitivity. Each 8$-$10 hour session observed 7 or 8 pointings, cycling among them for improved $uv$ coverage giving roughly one hour on source for each pointing. Sixty-one antennas were used for the observations contributing to the mosaic used in this work.

Calibration and imaging of the pointing centres were as described in \cite{2020ApJ...888...61M}. A linear mosaic of the pointing images was made as described in \cite{2021A&A...651A..85B}. The resulting mosaic image has a synthesised beam of 8\arcsec$\times$8\arcsec\ and the \ac{rms} noise is 56\mujybm. We utilized only a total-intensity image throughout the paper. 

\subsection{Parkes Observatory}
\label{subsec:parkes}

Parkes radio telescope, also known as Murriyang, is a 64-m paraboloid single-dish antenna. On 10\,September 2019, as part of project P1019 \citep{parkes}, Potoroo was observed using the \ac{UWL} receiver in conjunction with the Medusa backend in the pulsar search mode. The data were recorded with 2-bit sampling every 64\,$\mu$s in each of the 0.125\,MHz wide frequency channels, resulting in a total of 26,624 channels covering from 704 to 4032\,MHz. The observations were carried out for a total integration time of 4\,hours and only the total intensity was recorded. We performed a periodicity search using the pulsar software package PRESTO\footnote{\url{https://github.com/scottransom/presto}} \citep{ransom01} for every 512\,MHz of bandwidth and for a \ac{DM} range from 0 to 2000\,\dmunits. At frequencies above 3\,GHz, we identified a pulsar candidate.

Follow-up observations were conducted on 12\,July and 12\,October\,2022, using the coherently de-dispersed search mode. In this mode, data were recorded with 2-bit sampling every 64\,$\mu$s in each of the 1\,MHz wide frequency channels (3328 channels across the entire band with Medusa). We recorded full Stokes information for both follow-up observations. To calibrate these observations, we observed a pulsed noise signal injected into the signal path before the first-stage low-noise amplifiers before each observation.

The pulsar candidate, \psr, was detected and confirmed in both observations with high significance. We determined the apparent spin period and \ac{DM} for each observing epoch and folded the data using the \texttt{DSPSR}~\citep{2011PASA...28....1V} software package with a sub-integration length of 30\,s. Data affected by narrow-band and impulsive \ac{RFI} were manually excised using the \texttt{PSRCHIVE}~\citep{2004PASA...21..302H} software package. Polarisation and absolute flux calibrations of these search mode observations were carried out following steps described in \cite{2019ApJ...874L..14D}. Each observation was then averaged in time to create sub-integrations with a length of a few tens of minutes, and the pulse time of arrivals was measured for each integration using \texttt{PSRCHIVE}. The spin period of \psr\ at each observing epoch was then measured using the \texttt{Tempo2} software package~\citep{2006MNRAS.369..655H}. 

\subsection{Chandra X-Ray Observatory}
\label{subsec:chandra}

The extended X-ray source of Potoroo was serendipitously discovered during the \chandra\ survey of the Norma region of the Galactic spiral arms. The source was observed on 13\,June\,2011 with two exposures, 19.31\,ks and 19.01\,ks, with \acp{ObsID} 12519 and 12520, respectively. The data were taken with the \ac{ACIS-I} operating in Very Faint (VFAINT) timed exposure mode. The \ac{ACIS-I} covers a 16.9\arcmin$\times$16.9\arcmin\ \ac{FoV} and has the on-axis spatial resolution less than 0.5\arcsec, which increases off-axis. Both observations had Potoroo positioned off-axis, less so in the \ac{ObsID}\,12519, with Potoroo located $\sim$\,3.8\arcmin\ away from the optical axis, compared to $\sim$\,8.4\arcmin\ for the \ac{ObsID}\,12520. 
 
The energy scale of the CCD chip is calibrated over the range of approximately $0.3-10$\,keV. The \ac{ACIS-I} has front-side-illuminated CCDs where the instrumental background dominates the spectrum for energies below 0.5\,keV and above $7-8$\,keV \citep{2003ApJ...591..891B}. The time resolution of the CCD chip, determined by the readout time, was 3.2\,s.

We downloaded the observations from the \ac{CDA} and reduced the data using the \ac{CIAO}\footnote{\url{https://cxc.cfa.harvard.edu/ciao}} version 4.15 \citep{2006SPIE.6270E..1VF} with the most recently available \ac{CALDB} version 4.10.2. The standard \textsc{chandra\_repro} tool was performed to reprocess both sets of observations and \textsc{mdcopy} to filter the event files from background noise. Since Potoroo has an extended structure, reducing the background noise makes fine structures more prominent. To increase the signal-to-noise ratio, utilizing the \textsc{merge\_obs} tool, the observations were reprojected and combined to create a merged event file and exposure-corrected images. For the object's offset from the optical axis, we calculated the resolution of the merged image to be 2.4\arcsec$\times$2.4\arcsec\ using the default \textsc{psfmerge} setting, based on the 90\% point-spread function. Since no source events below 2\,keV were detected, the final image covers the energy range of $2-8$\,keV. 

We present the archival \chandra\ data only for the purpose of comparing the X-ray imaging results to the new radio observations of Potoroo. The exposure-corrected and merged image is additionally smoothed with a two-dimensional Gaussian kernel and variance of $\sigma$\,=\,2.5\arcsec\ to reduce the effect of statistical fluctuation (Figure\,\ref{fig:collage}).

\subsection{Wide-Field Infrared Survey Explorer}
\label{subsec:wise}

The \ac{WISE} satellite \citep{2010AJ....140.1868W} performed all-sky mid-infrared surveys in four photometric bands centered on 3.4\,$\mu$m, 4.6\,$\mu$m, 12\,$\mu$m and 22\,$\mu$m with an angular resolution of 6.1\arcsec, 6.4\arcsec, 6.5\arcsec\ and 12.0\arcsec, respectively. The higher two bands are sensitive to stars, while 12\,$\mu$m and 22\,$\mu$m maps contain a wealth of gas and dust information. With high sensitivity, the satellite was able to detect over 8000 \HII~regions in the Galactic plane.  

We used the \ac{WISE} emission maps at 12\,$\mu$m and 22\,$\mu$m and the \HII~regions catalogue \citep{2014ApJS..212....1A} to search for the Potoroo parent \ac{SNR}. Synchrotron non-thermal emission, such as from shell \acp{SNR}, should not correlate with the mid-infrared thermal emission, which makes a powerful tool for distinguishing \acp{SNR} from \HII~regions. However, in complex parts of the \ac{ISM}, such as the Potoroo's surrounding environment, other emissions may be present in the line of sight that could lead to the misclassification of objects.

\section{Results of Potoroo Study}
\label{sec:results}

\begin{figure*}
\centering
\includegraphics[trim= 0 0 0 0 ,width=\textwidth]{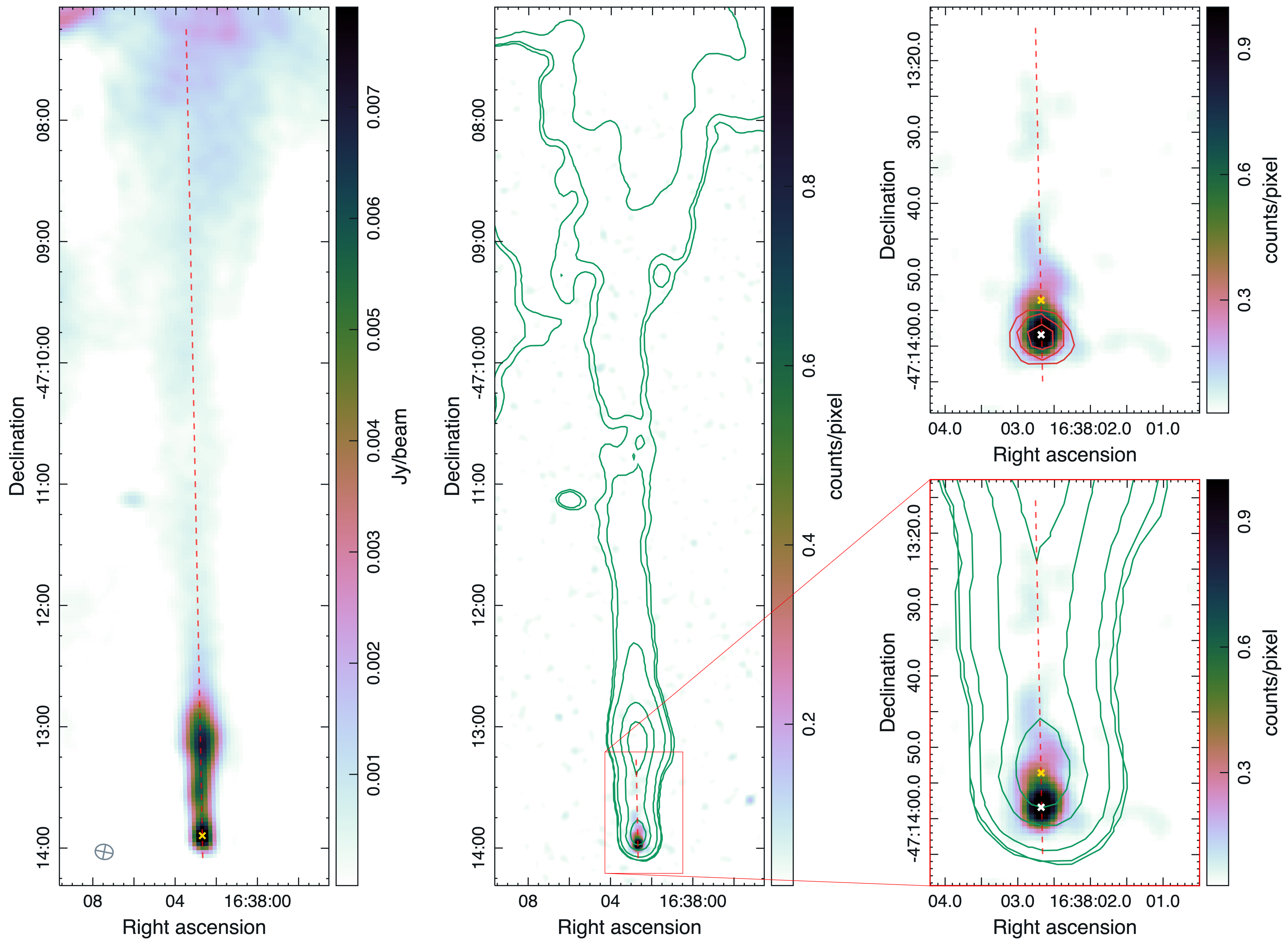}
\caption{Radio and X-ray images of Potoroo. The left panel presents the \ac{ASKAP} total intensity image at 1368\,MHz, with the beam size of 8.8\arcsec$\times$7.4\arcsec shown in the bottom left corner. In the middle panel, the \chandra\ image within the energy range of $2-8$\,keV is smoothed with a two-dimensional Gaussian where $\sigma$ = 2.5\arcsec. The green contours correspond to \ac{ASKAP}'s 1368\,MHz Stokes\,$\rm I$ image at the following levels: 0.2, 0.3, 0.8, 3, 5 mJy\,beam$^{-1}$. The red box highlights the area of the zoomed-in \chandra\ image in the right panels, where the bottom right image has the same \ac{ASKAP} contours as the middle image and the top right image is overlaid with red contours from \ac{ASKAP} 944\,MHz image of circular polarisation. The significances of the red contours are 3, 4 and 5$\sigma$, where 1$\sigma$=24\,$\mu$Jy\,beam$^{-1}$. The white ``x'' marks the X-ray peak, while the yellow ``x'' marks the radio peak. The red dashed lines denote the axis of symmetry of Potoroo.}
\label{fig:collage}
\end{figure*}

Combining \ac{ASKAP} radio-continuum data with the infrared maps from the \ac{WISE} survey (Figure\,\ref{fig:rgb}) and \chandra\ X-ray image (Figure~\ref{fig:collage}), we can confidently classify Potoroo as a bow-shock \ac{PWN} positioned within the plane of our galaxy. 

\begin{figure*}
\centering
\includegraphics[width=\textwidth]{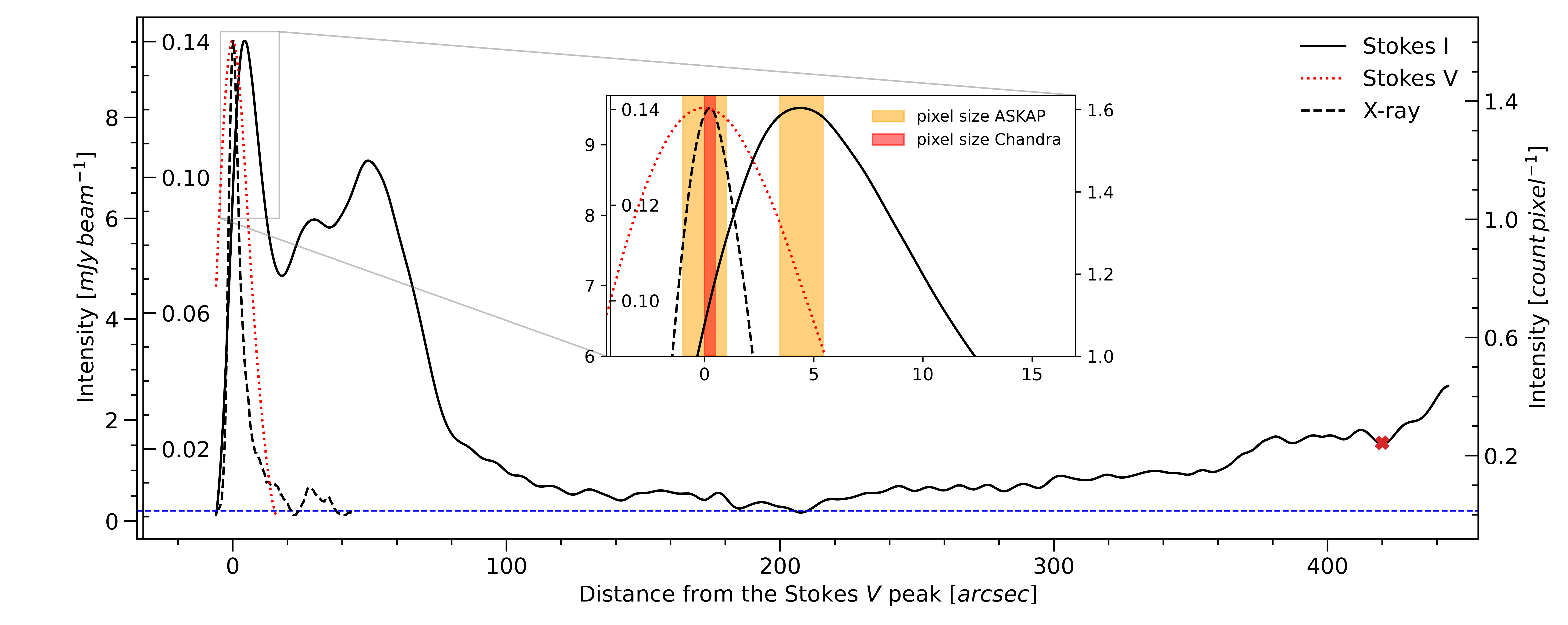}
\caption{ Profiles of Potoroo along the axis of symmetry as a function of the distance from the Stokes\,$\rm V$ peak. The Stokes\,$\rm I$ profile at 1368\,MHz is represented with a black solid line, while the Stokes\,$\rm V$ profile at 944\,MHz is represented with the red dotted line. Both profiles are given in \lbrack mJy\,beam$^{-1}$\rbrack\ units and correspond to the left y-axes, respectively. The X-ray intensity profile in the 2 – 8 keV energy range is shown by the black dashed line and corresponds to the right y-axis. The grey box indicates the region of the zoomed-in plot showing the peaks of the Stokes\,$\rm V$, X-ray and Stokes\,$\rm I$ profiles, listed in order of appearance. The orange and red shaded boxes represent the pixel increment of the \ac{ASKAP} and \chandra\ data. The horisontal dashed blue line denotes the noise level, and the red "x" marks the end of the tail studied in this paper.}
\label{fig:profiles}
\end{figure*}

The distance to Potoroo was previously estimated by \cite{2014ApJ...787..129J} based on foreground \HI\ absorption, suggesting a lower bound of 10\,kpc. This places Potoroo in the Norma spiral arm on the far side of the Galaxy (Norma\,II region). We also consider another potential distance of 7.6\,kpc, calculated from the Potoroo pulsar's \ac{DM}, as discussed in Section\,\ref{subsec:distance}.

\cite{2014ApJ...787..129J} reported the X-ray source position coordinates as RA(J2000)\,= \,16$^{\rm h}$38$^{\rm m}$02.7$^{\rm s}$, Dec(J2000)\,= $-$47\D13\arcmin58.4\arcsec, peaking at an energy maximum of $\approx$\,5\,keV. The radio position of Potoroo, estimated at the peak of radio brightness, has coordinates: RA(J2000)\,= \,16$^{\rm h}$38$^{\rm m}$02.6$^{\rm s}$ and Dec(J2000)\,=\,$-$47\D13\arcmin53.8\arcsec. We measured the flux densities for the entire object using the techniques described in \citet{2022MNRAS.512..265F}. After convolving images to the same resolution as the lowest-resolution image, we determined total radio flux densities for the carefully selected 3$\sigma$-source region, accounting for the local background. These measurements, with a flux density measurement scale error of approximately 10\% of the overall flux density, are listed in Table\,\ref{tab:flux}.  

\begin{table}
\begin{adjustbox}{width=\columnwidth,center}
\caption{Peak flux densities $S_{\rm peak}$ and integrated flux densities $S_{\nu}$ of Potoroo at three frequencies. The average polarised intensity $PI$ and fractional polarisation $FP$, as well as the peak of fractional polarisation $FP_{\rm peak}$ are quantified only for 1368\,MHz \ac{ASKAP} data.}
\label{tab:flux}
\begin{tabular}{c c c c c c} 
\toprule
\makecell{$\nu$ \\ \lbrack MHz \rbrack} & \makecell{$S_{\rm peak}$ \\ \lbrack mJy\,beam$^{-1}$ \rbrack} & \makecell{$S_{\nu}$ \\ \lbrack mJy \rbrack} & \makecell{$PI$\footnote{Integrated values are measured for the selected regions that match the 3$\sigma$ source significance where
$\sigma_{1368MHz}$\,=\,68\mujybm. The polygonal contours are created with a \textsc{polygon} and the values are calculated with the \textsc{statistics} package from \ac{CARTA} \citep{carta}.\label{region_note}} \\ \lbrack mJy \rbrack} & \makecell{$FP$\footref{region_note} \\ \lbrack\%\rbrack} & \makecell{$FP_{\rm peak}$ \\ \lbrack\%\rbrack}\\ 
\hline
\hline
 944 & 15.58 & 389\,$\pm$\,40 & $-$ & $-$ & $-$ \\ 
1284 & 10.91 & 293\,$\pm$\,30 & $-$ & $-$ & $-$ \\
1368 & ~9.51 & 238\,$\pm$\,25 & 5.90\,$\pm$\,0.33 & 6.84\,$\pm$\,0.45 & 24
\end{tabular}
\end{adjustbox}
\end{table}

The positional difference between the X-ray and radio peaks is 4.7\arcsec, as shown at the bottom right panel of Figure\,\ref{fig:collage} and Figure\,\ref{fig:profiles}. Given the uncertainty for the \chandra\ position of 0.36\arcsec\ \citep{2014ApJ...796..105F} and the astrometric precision for ASKAP of about 0.6\arcsec\ \citep{2022MNRAS.512.6104G} for both RA and Dec, the offset must be real. Such an offset has already been seen in other bow-shock \ac{PWN}e like the Lighthouse nebula \citep{2014A&A...562A.122P} and G319.9$-$0.7 \citep{2010ApJ...712..596N}.

Investigating the 944\,MHz ASKAP circular polarisation (Stokes\,$\rm V$) image of Potoroo, we observe a similar positional difference between the Stokes\,$\rm V$ peak and the peak of the total intensity image (Stokes\,$\rm I$). Additionally, the Stokes\,$\rm V$ peak aligns well with the peak of the X-ray source. Both comparisons are shown in Figure\,\ref{fig:collage},\,right and Figure\,\ref{fig:profiles}. The circularly polarised emission is detected with a significance greater than 5$\sigma$, with a local rms noise of about 24\mujybm. The fractional circular polarisation, expressed as the V/I ratio, is slightly higher than 1\,$\%$. Despite the weak circular polarisation, it cannot be attributed to leakage from Stokes\,$\rm I$, as the peak of the circularly polarised emission is offset from the total intensity peak. Moreover, the leakage across the entire ASKAP \ac{FoV} is typically around 0.7\,$\%$.

The circularly polarised emission is almost unique to stellar objects \citep{2015MNRAS.449.3223D,2018MNRAS.478.2835L}, although it is not universal among them \citep{1998MNRAS.301..235G}. However, circularly polarised sources that lack clear optical or infrared association are strong pulsar candidates. Motivated by this, together with the far distance and potentially high dispersion of Potoroo's pulsar, we reprocessed the archival Parkes data, focusing on the high-frequency band, rather than the L-band typically used for pulsar detection.  We identified and confirmed a periodic signal from the pulsar at a \ac{DM} of 1553\,\dmunits\ with a spin period of 65.74\,ms. For more details on the \psr\ detection and its properties, see Sections\,\ref{subsec:pulsar} and \ref{subsec:pulsar_discussion}. 

\subsection{Potoroo Morphology}
\label{subsec:morphology}

The morphology of \ac{PWN}e is mainly determined by the properties of the ambient medium and the characteristics of the pulsar that powers the \ac{PWN}. The striking cometary structure of Potoroo is the typical shape of a \ac{PWN} whose pulsar is moving supersonically through the ambient medium. In this case, the ram pressure exerted by the oncoming medium confines and channels the pulsar wind in the opposite direction to the pulsar motion, as shown in Figure\,\ref{fig:collage}, left.

Visual inspection of the radio total intensity images reveals an extended morphological structure of Potoroo, characterized by two distinct components:  a compact and bright head followed by a highly elongated tail. The X-ray source has a similar structure but on a different length scale. In addition to the mismatch of the radio and X-ray intensity peaks, Figure\,\ref{fig:profiles} shows that the radio emission extends beyond the X-ray emission. This extension is attributed to the synchrotron ageing effect of relativistic electrons, where low-energy electrons lose energy at a slower rate, resulting in longer lifetimes that allow them to travel farther from the pulsar.


Potoroo is roughly conical in shape. The radio source extends approximately 7.2\arcmin\ along the \ac{PA} of 0\D, equivalent to a physical size of 21\,pc considering the distance of 10\,kpc. After a dip in intensity behind the head, the brightness increases, peaking at a distance of about 0.8\arcmin. Following this peak, it rapidly dims and remains relatively unchanged. The bright structure (hereafter the main body) covers the first 1.9\arcmin\ of the object. The tail has a hazy termination due to the complexity of the surrounding medium. Our estimation of the source end is based on the morphological appearance and loss of the tail shape (Figure\,\ref{fig:collage}). Additionally, this estimation aligns with a significant increment in the slope of the brightness profile influenced by ambient emission (Figure\,\ref{fig:profiles}). However, we could not exclude that the source is longer than that. In contrast, the X-ray tail brightness fades to background levels at approximately 0.67\arcmin\ (2\,pc for the distance of 10\,kpc) from the pulsar. 

The radio shape of the source is narrow overall, ranging from 18\arcsec$-$\,24\arcsec, until it abruptly widens by a factor of 3 at a distance of about 4.4\arcmin. The rapid change in tail width is likely due to an ambient density discontinuity. As the ambient density increases along a pulsar path, the ram pressure also increases, causing the \ac{PWN} to adopt a thinner shape \citep{2017MNRAS.464.3297Y}. However, if the ambient density remains unchanged over a sufficient distance, the tail flattens again, as observed in other parts of Potoroo.

The \chandra\ observation of Potoroo has also detected a diffuse feature perpendicular to the \ac{PWN} axis, extending approximately 19\arcsec\ (see Figure\,\ref{fig:collage}, right). However, interpreting this feature is challenging due to limited statistical data. Likewise, the radio emission data do not reveal any visible counterpart, similar to other objects with more prominent X-ray misaligned outflows, such as the Lighthouse \citep{2014A&A...562A.122P,2022arXiv221203952K} or Guitar nebula \citep{2007A&A...467.1209H,2022ApJ...939...70D}.

\subsection{Radio Continuum Spectrum of Potoroo}
\label{subsec:spectral_analysis}

\ac{PWN}e emit synchrotron radiation detectable across the electromagnetic spectrum, from radio to beyond the X-ray frequencies. In the radio band, the synchrotron emission is characterised by a power law distribution of flux density, expressed as S$_{\nu}\propto\nu^{\alpha}$. \ac{PWN}e typically have flat spectra with a spectral index in the range of $-$0.3\,$<\alpha<$\,0 \citep{1988ARA&A..26..295W, 2017ASSL..446....1K}. In rare cases, \ac{PWN}e spectra can be steeper, with the index as low as $-$0.7, as observed, for instance, in Dragonfly \citep[$\alpha=-0.74,$][]{2023ApJ...942..100J}. 

\begin{figure}[hbt!]
\centering
\includegraphics[width=0.75\columnwidth]{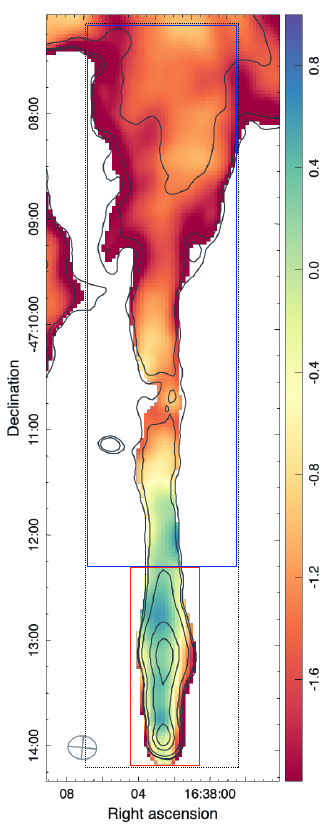}
\caption{Spectral index map created using the \ac{ASKAP} total intensity images at 944\,MHz and 1368\,MHz, and the MeerKAT image at 1284\,MHz. The map is overlaid with the same radio contours as presented in Figure\,\ref{fig:collage}, middle panel. In the bottom left corner, the synthesized beam size, 16.3\arcsec$\times$13.3\arcsec, is given by a grey ellipse. The boxes indicate specific regions for which spectral indices are calculated: the main body (red box), the diffuse region (blue box) and the entire object (black dotted box).}
\label{fig:spectral_index}
\end{figure}

A spectral index map of Potoroo was created using Stokes\,$\rm I$ images at 944, 1284 and 1368\,MHz with the MIRIAD software package \citep{miriad_1995ASPC...77..433S}. To maintain pixel consistency across all bands, the \ac{ASKAP} images were first regridded to match the finest pixel grid, 1.5\arcsec$\times$1.5\arcsec, of the MeerKAT 1284\,MHz image. Then, all images were convolved to the lowest resolution of the dataset, corresponding to the resolution of the \ac{ASKAP} 944\,MHz image with a beam size of 16.3\arcsec$\times$\,13.3\arcsec. The local 1$\sigma$ \ac{rms} noise levels of the individual images were measured using the \textsc{rms statistics} tool in the \ac{CARTA} software. These measurements were performed on carefully selected regions, excluding all obvious sources. The convolved images were combined, and a spectrum was fitted using a simple weighted linear regression algorithm. The algorithm modelled pixels with flux densities more significant than 5$\sigma$ \ac{rms} mask thresholds. The flux densities of the same pixel for all bands were fitted, and the resulting slope of the best line fit was stored in a new image. The distribution of these slopes forms the spectral index map of Potoroo, as shown in Figure\,\ref{fig:spectral_index}. Pixels with flux densities less than the 5$\sigma$ threshold at least one frequency have default NaN values, do not contribute to future calculations, and are represented as a white background.

The average spectral index measured across Potoroo is $-$1.26, with a standard deviation of 0.04. The measurement was obtained using the spectral index map for the region marked with the black dotted box in Figure\,\ref{fig:spectral_index}. The average value is calculated using the \textsc{mean statistics} tool in the \ac{CARTA}.

The overall spectral index we obtained is significantly lower than the typical values for \ac{PWN}e, but it aligns with our expectations for the entire object, covering both the bright and diffuse parts. The steepest section of the Potoroo spectral map corresponds to the large, diffuse area of the tail (marked with the blue box in Figure\,\ref{fig:spectral_index}), with an average spectral index of $-$1.42\,$\pm$\,0.04. Consequently, the overall spectral index steepens considerably when the diffuse part is included. However,  if we only consider the main body (marked with the red box), the calculated average spectral index is $-$0.21$\pm$0.02. This value is consistent with most \ac{PWN}e, including the Crab \citep[$\alpha=-0.27$,][]{crab}, G319.9$-$0.7 \cite[$\alpha=-0.26$,][]{2010ApJ...712..596N} and the Boomerang \cite[$\alpha=-0.11$,][]{2006ApJ...638..225K}. In Section\,\ref{subsec:break}, we discuss these radio spectrum results and provide a comparison with the X-ray spectrum.

\subsection{Polarisation Analysis}
\label{subsec:polarisation}

The flat spectral index of \ac{PWN}e is similar to optically thin thermal emission from \HII~regions. However, unlike thermal emission, synchrotron radiation in \ac{PWN}e is linearly polarised, with the degree of polarisation ranging from a few per cent to more than 30\% \citep[e.g.,][submitted]{2022arXiv220811026M}. Typically, the magnetic field within \ac{PWN}e is assumed to be toroidal \citep{2003A&A...404..939V, 2017SSRv..207..137P}. To investigate the internal magnetic field of \ac{PWN}e, we need to separate internal Faraday rotation from foreground effects. This process usually requires using three frequencies for accurate measurements.

\begin{figure*}[hbt!]
 \begin{center}
 \includegraphics[width=0.33\textwidth]{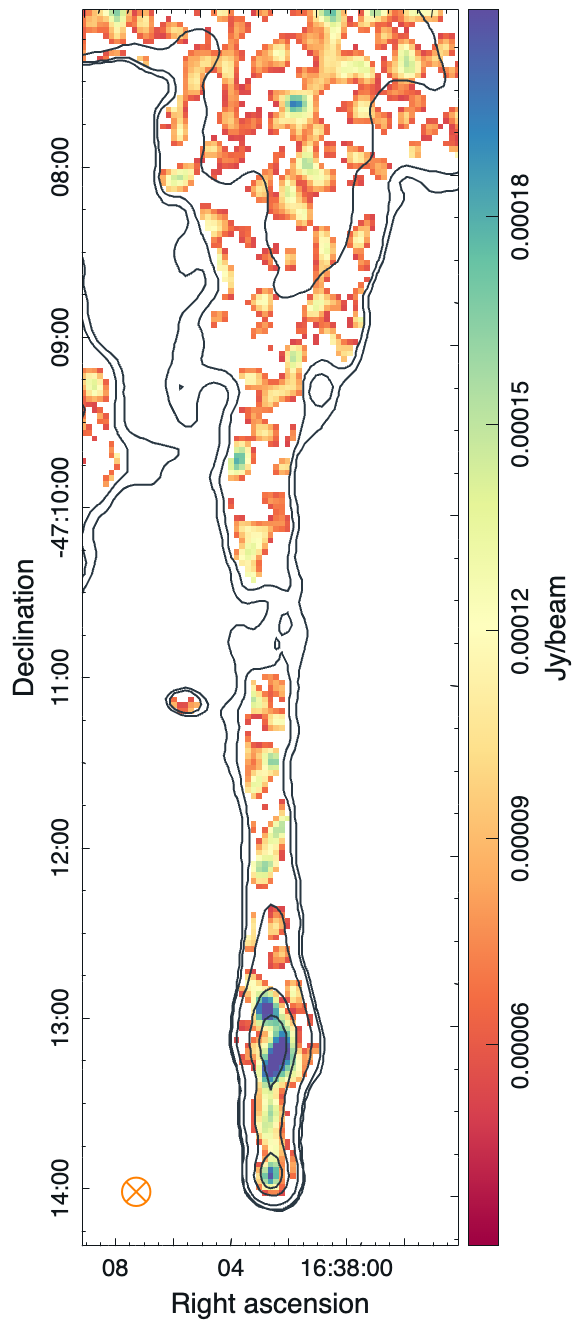}
 \includegraphics[width=0.33\textwidth]{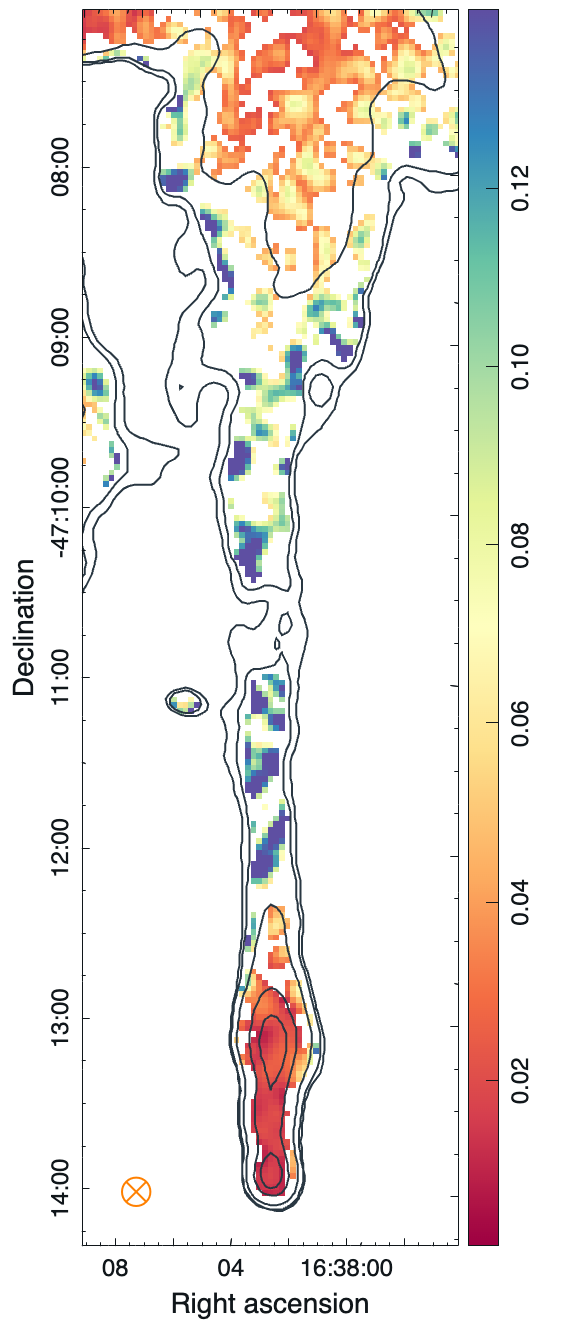}
 \includegraphics[width=0.33\textwidth]{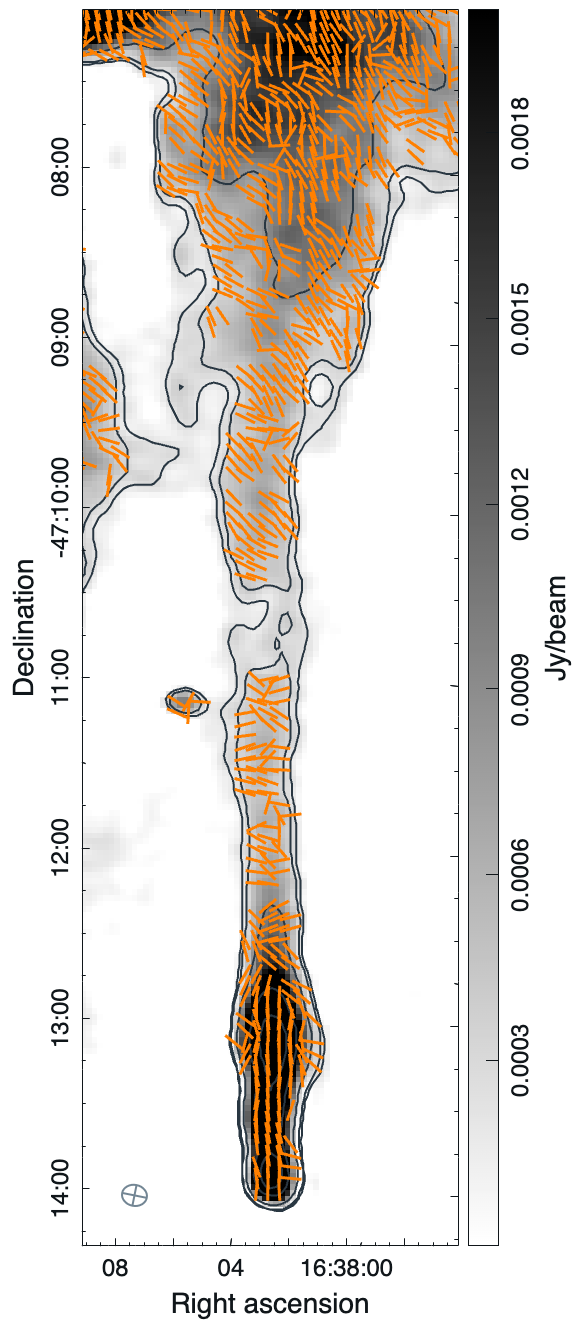}
 \end{center}
 \caption{Polarisation intensity (left) and fractional polarisation (middle) maps of Potoroo at 1368\,MHz are shown. The orange ellipses in the bottom left corner represent the synthesised beam with the size of 10\arcsec$\times$10\arcsec. Polarisation vectors in the observed electric field direction overlay the grey-scale Stokes\,$\rm I$ image (right). The plotted vectors are of equal length and approximately half the beam size apart. All images have the same superimposed contours as used in the middle panel of Figure\,\ref{fig:collage}.}
 \label{fig:polarisation}
\end{figure*}

Polarisation analysis of Potoroo was performed using \ac{ASKAP} 1368\,MHz data only, as no Stokes\,$\rm Q$ and $\rm U$ data were available at any other frequency. Due to low signal-to-noise ratios, the full resolution polarisation Stokes\,$\rm Q$ and $\rm U$ images were convolved to a common resolution of 10\arcsec$\times$10\arcsec. The polarisation intensity, fractional polarisation and polarisation angle maps, along with their errors, were calculated using the standard MIRIAD task \textsc{impol}. Only pixels exceeding 5$\sigma$ noise mask were considered. The resulting maps are presented in Figure\,\ref{fig:polarisation}, and the integrated polarisation intensity, fractional polarisation, and peak fractional polarisation are summarised in Table\,\ref{tab:flux}. 

The left panel of Figure\,\ref{fig:polarisation} shows that the linear polarised emission is strongest toward the head and bright features of Potoroo's tail. It is not surprising that the polarised intensity correlates with the total intensity. There seems to be patches of polarised emission outside of the bright region of the \ac{PWN}, associated with the diffuse part of the tail. However, this emission is too faint to be properly quantified. 

The middle and right panels of Figure\,\ref{fig:polarisation} display the fractional polarisation map and polarisation vectors in the observed electric field direction, overlaying the Stokes\,$\rm I$ image. The vector lengths are left unscaled for clarity. Our analysis reveals that the polarisation fraction for the entire Potoroo object can reach up to 24\%, with an average value below 7\%. Focusing solely on the main body, the area with a high signal-to-noise ratio, we observe weak polarisation of approximately 2\%, with a peak of about 6\% and highly ordered polarisation vectors. The overall lower fractional polarisation could be attributed to internal Faraday rotation, which can cause significant depolarisation at lower frequencies \citep[e.g.,][]{2006ApJ...638..225K}. 

The polarisation E-vectors observed in the vicinity of the place of origin exhibit a parallel orientation relative to the Potoroo axis, indicating magnetic field vectors running in the tangential direction. This magnetic field geometry is similar to the findings in G319.9$-$0.7 \citep{2010ApJ...712..596N} and Boomerang \citep{2006ApJ...638..225K}. In contrast, for the Mouse \citep{2005AdSpR..35.1129Y} and G315.9$-$0.0 \citep[the Frying Pan \ac{SNR}, ][]{2012ApJ...746..105N}, the magnetic field shows a parallel alignment with the tail. The polarisation vectors of Potoroo switch orientation with the distance from the pulsar and become disordered, but some radial tendencies can be seen. This change is also detected in G319.9$-$0.7 but not in G315.9$-$0.0.

It is important to note that polarisation measurements should be considered as indications only, as we used Stokes\,$\rm Q$ and $\rm U$ maps at one frequency. Without additional data, a precise determination of Faraday rotation and magnetic field properties is not possible. The complete polarisation study will be presented in a subsequent paper.

\subsection{Detection of PSR~J1638--4713}
\label{subsec:pulsar}

\psr\ has a spin period of 65.74\,ms and the second-highest \ac{DM} of all known radio pulsars. Its \ac{DM} of 1553\,\dmunits\ is only exceded by the Galactic Centre magnetar~\citep[PSR~J1745$-$2900,][]{2013Natur.501..391E}. According to the YMW16\footnote{\url{https://www.atnf.csiro.au/research/pulsar/ymw16}} electron density model~\citep{2017ApJ...835...29Y}, this gives a \ac{DM} distance of $\sim7.6$\,kpc, although this could be highly uncertain considering the complexity in the Galactic plane.

\begin{figure}[hbt!]
\centering
\includegraphics[width=\textwidth]{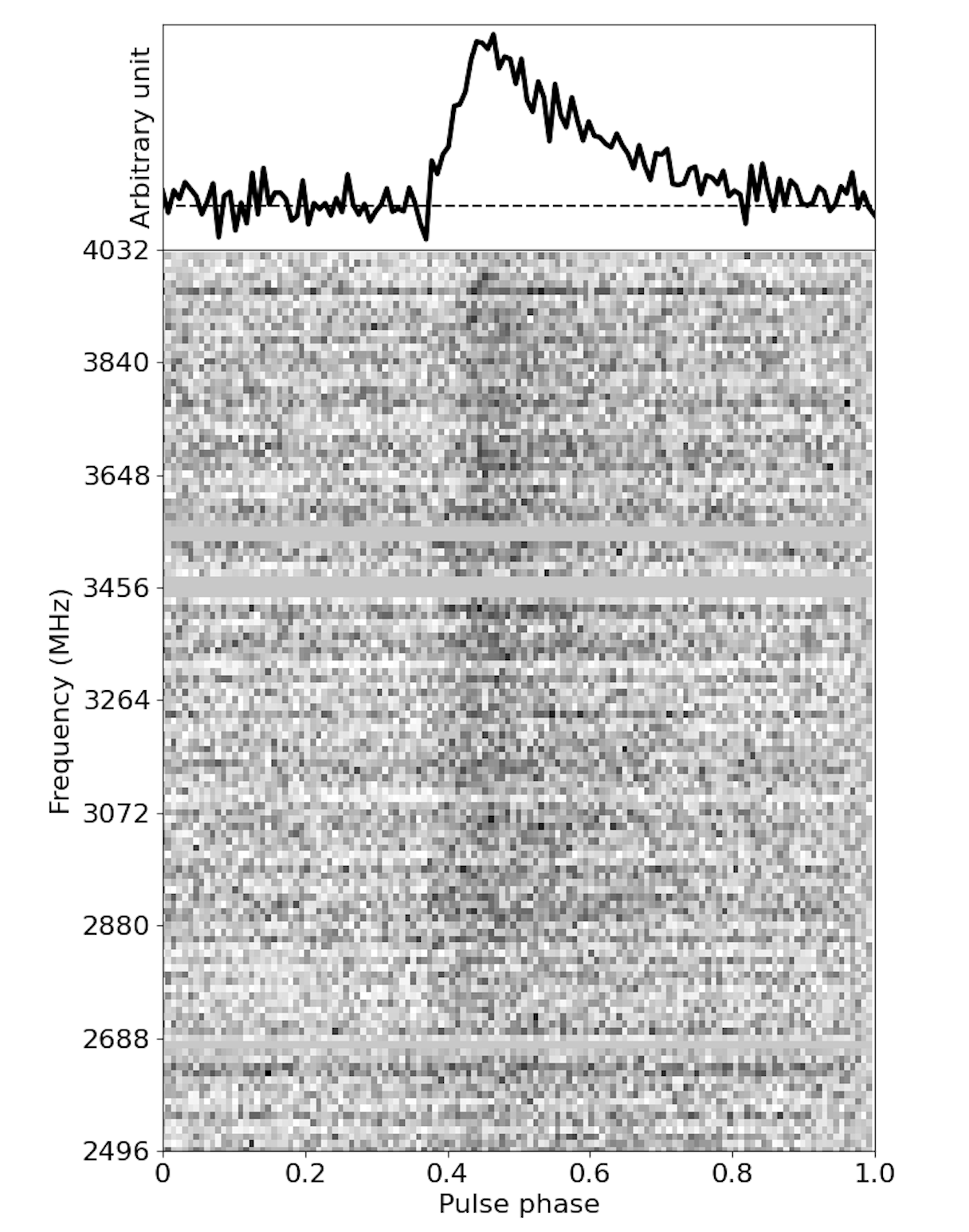}
\caption{The pulse profile (top panel) and frequency spectrum (bottom panel) of \psr, the 65.74\,ms pulsar discovered in the Potoroo \ac{PWN} with the Parkes UWL at 3\,GHz. The pulse profile has been corrected for the measured \ac{DM} of 1553\,\dmunits. Greyed out horizontal bands, such as at a frequency of 3456\,MHz,
represent data flagged to remove \ac{RFI}.}
\label{fig:pulse_profile}
\end{figure}

The large \ac{DM} results in the strong scattering we observed in \psr\ and its pulse is completely scattered below $\sim$2500\,MHz (Figure\,\ref{fig:pulse_profile}). This explains why this pulsar was not discovered by previous pulsar surveys. We measured the scattering time-scale to be $6.2\pm0.7$\,ms at 3700\,MHz by fitting the pulse profile with a Gaussian intrinsic pulse profile convolved with an exponential tail~\citep[e.g.,][]{2022MNRAS.513.1794B}. Assuming that the scattering time-scale scales as $\nu^{-4}$, we estimated the scattering time-scale to be $\sim70$\,ms at 2\,GHz, which is longer than the pulsar spin period and explains its non-detection at lower frequencies. 

\begin{figure}[hbt!]
\centering
\includegraphics[width=\textwidth]{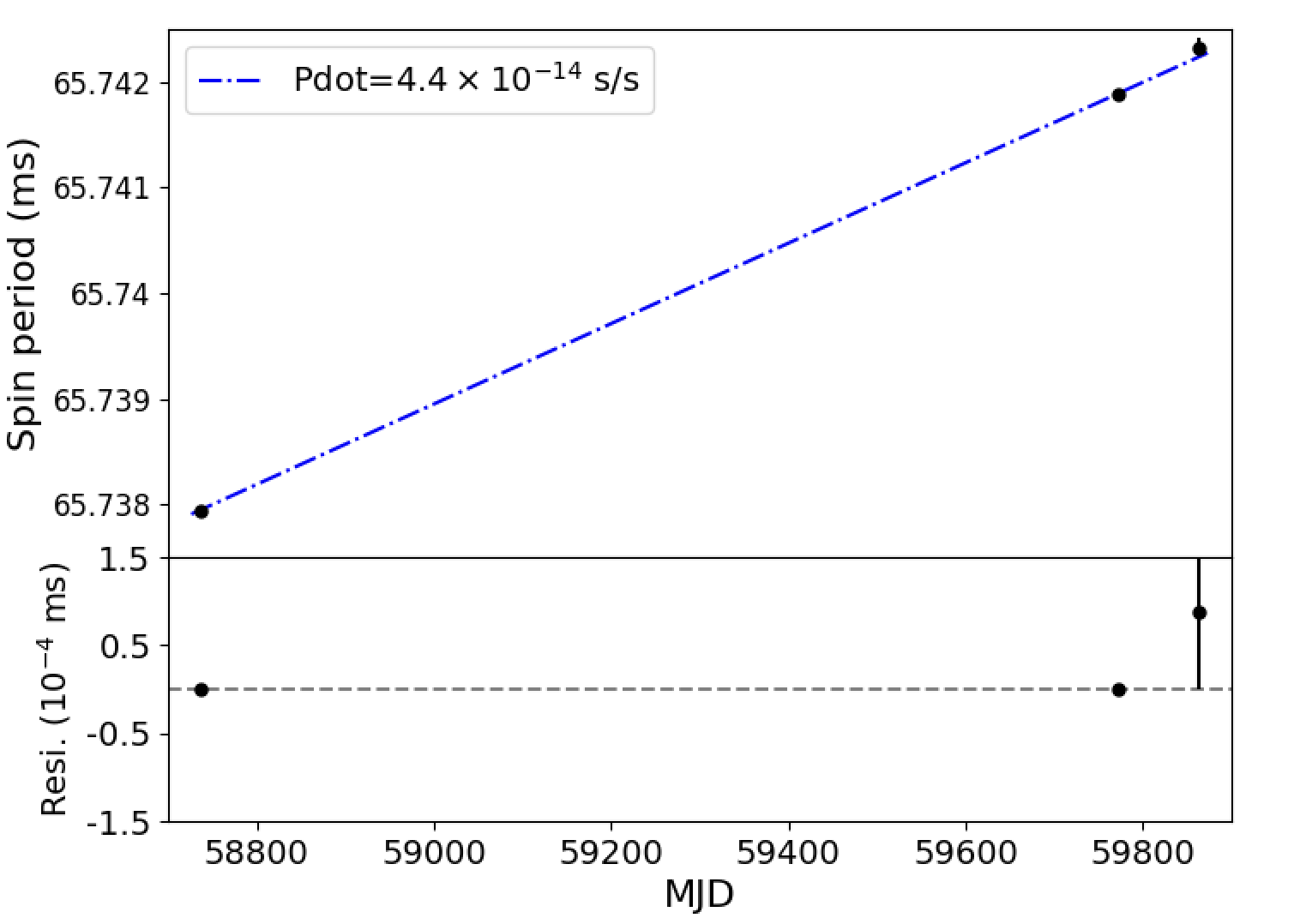}
\caption{Pulsar spin period as a function of time. The dot-dashed line shows a linear fit, which gives a measure of the spin-down, $\dot{P}=4.4\times$10$^{14}$s/s.}
\label{fig:spin_down}
\end{figure}

After averaging frequencies from 2496 to 4032\,MHz, we measured a pulsar flux density of $226\pm5\,\mu$Jy at 3264\,MHz. This is much smaller than the continuum flux density of Potoroo (see Table~\ref{tab:flux}) and suggests that the observed radio continuum emission is dominated by the \ac{PWN}. No polarised signal has been detected in the pulsed emission so far. Our current Parkes observations are not sensitive enough to detect the circularly polarised emission of $\sim0.05$\,mJy, detected in continuum images. 

In Figure\,\ref{fig:spin_down}, we show the measured spin period of the pulsar J1638$-$4713 as a function of time. We can clearly see the gradual spin-down of \psr\ over the course of $\sim3$\,yr. To estimate the spin-down rate of \psr, we fitted a constant spin-down rate (i.e. $\dot{P}$) to our measurements and obtained $\dot{P}=4.407(8)\times10^{-14}$. According to canonical pulsar spin-down models \citep[e.g.,][]{2004hpa..book.....L}, we estimated the characteristic pulsar age $\tau=P/2\dot{P}\approx24000$\,yr, the characteristic magnetic field $B_{\rm s}\approx1.7\times10^{12}$\,G and spin-down luminosity of $\dot{E}\approx6.1\times10^{36}$\,erg\,s$^{-1}$. These parameters indicate that \psr\ is a young pulsar with high spin-down luminosity, which is consistent with our other observations about Potoroo. 

\subsection{Unknown Origin of Potoroo}
\label{subsec:snr}

Pulsars are expected to originate from the explosion of a massive star when its core undergoes a rapid implosion. The explosion can give the neutron star a significant kick, causing it to move away from its birthplace at high speeds \citep{2004cetd.conf..276L}. We often observe a pulsar's trailing emissions pointing back to its origin. However, in the case of Potoroo, we were not able to associate the \ac{PWN} with any known \ac{SNR} or detect any sign of a remnant nebula from the explosion of the progenitor star. We also considered the possibility of an off-centre explosion and asymmetries in the surrounding medium. Nevertheless, this is not a unique case as similar examples, \ac{PWN}e without clear origins, have been observed, e.g. G319.9-0.7 \citep{2010ApJ...712..596N} and the Mouse \citep{2018ApJ...861....5K}. 

Figure\,\ref{fig:rgb} displays the mid-infrared view from \ac{WISE} using the \ac{ISM}-sensitive bands (12\,$\mu$m in green and 22\,$\mu$m in blue image layers), alongside the radio continuum view (red layer) of the Galactic plane region in the vicinity of Potoroo. Radio emission from \acp{SNR} is primarily synchrotron and should lack clear mid-infrared counterparts. However, a large emission lump located behind the Potoroo trail shows a strong correlation between infrared and radio, indicating its thermal nature originating from \HII~regions. Moreover, the superposition of the known \HII~regions (green circles) confirms the previous. No clear evidence has been found for a shell marking the location of the expanding \ac{SNR} shockwave outside the \ac{PWN}.

The absence of an apparent parent \ac{SNR} could be explained by Potoroo's position in the far Norma arm, known for its abundance of dense molecular clouds, gas, and dust \citep{2001ApJ...547..792D}. These components could potentially obscure emissions coming from behind \citep[in prep.]{Ball}. In Section\,\ref{subsec:distance}, we discuss further the possible origins of Potoroo. 

\section{Discussion}
\label{sec:discussion}

\cite{2014ApJ...787..129J} detected extended X-ray emission in the Norma arm using \chandra\ survey data and tentatively classified it as a bow-shock tail \ac{PWN}. In addition to X-ray, the authors used radio data of the same field with the highest available resolution at that time, 60\arcsec$\times$45\arcsec\ for the particular \ac{MGPS-2} mosaic image at 843\,MHz, and found a radio trail coinciding with the X-ray peak position \citep[see][Figure 5]{2014ApJ...787..129J}. No radio structure could be resolved. With the recent development of radio telescopes which offer superior sensitivity and high-resolution capabilities, we are able to observe the complex structure of Potoroo in greater detail. Based on the results from the previous section, we can confidently classify Potoroo as a member of a small yet growing group of supersonic pulsars driving a bow-shock through the ambient medium. 

\subsection{How Far has Potoroo Travelled?}
\label{subsec:distance} 

The distance to Potoroo was determined using two different techniques. The first technique involved measuring atomic hydrogen absorption from the spectral fit of the entire X-ray source and comparing it to other objects in the same area. This method suggests that Potoroo is located in a far Norma~II region, with a minimum distance of 10\,kpc. The second technique involved using the YMW16 model which calculates the density of Galactic electrons and estimates the distance based on the pulsar’s \ac{DM}. The YMW16 model yielded a distance of 7.6\,kpc with an uncertainty of 20\%~\citep{2017ApJ...835...29Y}. However, it's important to note that our understanding of the distribution of free electrons weakens at large distances, particularly near the Galactic centre. The actual distance to Potoroo may have a larger error than the estimated 20\%.  

\cite{2007A&A...468..993K} calculated the kinematic distances to the Galactic spiral arms using a value for the Galactic centre of R$_{\odot}=7.6$\,kpc. Based on their calculations, the line of sight passes through the Norma~arm at distances of 4.9 and 11.5\,kpc with uncertainties of approximately $\pm1$\,kpc. Given this and considering that distance estimates based on pulsar \ac{DM} often underestimate the true value, we adopt a distance to Potoroo of 10\,kpc as a reasonable assumption and scale all distance-dependent quantities accordingly.

Potoroo is likely a mature nebula whose pulsar received enough of a natal kick to travel away from the centre of its parent \ac{SN} explosion, and the well-defined outer edges of the nebula suggest that Potoroo has interacted with the reverse shock of the \ac{SNR}. The duration for the reverse shock to reach maximum strength and dissipate varies depending on factors such as the density of the surrounding medium and the energy of the explosion. Previous studies suggest that, in denser environments, the reverse front can take around 10,000\,years to make its way back to the centre \citep{2001ApJ...563..806B,2001A&A...380..309V,2004AdSpR..33..475V}. We have used this estimation as the minimum age for Potoroo.

Although the Potoroo pulsar cannot be resolved in radio continuum images, the flat spectrum of the nebula close to the pulsar suggests the presence of a young, energetic electron population. The radio emission produced by these electrons indicates that the current energy input from the pulsar is significant for the total energy content of the nebula. According to \cite{2001ApJ...563..806B}, the energy input from a pulsar becomes negligible when the nebula's age is comparable to the characteristic age of the pulsar. Since the current characteristic age of the pulsar is approximately 23,000\,years and the energy input is still significant, we can infer that Potoroo is much younger than 24,000\,years.

Table\,\ref{tab:speed} presents various \ac{SNR} ages, ranging from 7.5\,kyr to 30\,kyr, that were used to estimate the transverse velocities of the pulsar. We also varied the travelled distances (one, two and three times the tail size) to examine the location of the parent \ac{SNR}. Assuming no inclination with respect to the plane of the sky, we calculated the velocities using the formula $V=2\,d\,t^{-1}\rm tan$$(\theta/2)$. As $\theta\ll1$, the relation simplifies to $V=2845.8\,d_{10}\,\theta_{\ast}\,t^{-1}_{\ast}$, where $d_{10}$ is the distance to Potoroo in units of 10\,kpc, $\theta_{\ast}$ corresponds to the angular length for a given travelled distance in arcmin, and $t_{\ast}$ stores the probing ages in k\,year.

\begin{table}
\caption{Natal kick velocities $V$ of the Potoroo pulsar derived as a function of \ac{SNR} age and travel lengths correspond to one, two and three times the tail sizes. The adopted distance to Potoroo is 10\,kpc. The analysis also includes the case of \ac{SNR} G338.1+0.4 at a distance of 6.2\,kpc. Reasonable values are highlighted in bold.}
\label{tab:speed}
\begin{tabular}{c c c c c | c} 
\toprule
\makecell{$\rm t$\\\lbrack k\,year\rbrack} & \makecell{} & \makecell{$V_{\rm tail}$\\\lbrack km\,s$^{-1}$\rbrack} & \makecell{$V_{\rm tail\times 2}$\\ \lbrack km\,s$^{-1}$\rbrack} & \makecell{$V_{\rm tail\times 3}$\\ \lbrack km\,s$^{-1}$\rbrack} & \makecell{$V_{\rm G338.1+0.4}$\\ \lbrack km\,s$^{-1}$\rbrack}\\
\hline
\hline
7.5 & & 2732 & 5464 & 8196 & 11763 \\ 
\textbf{10} & & \textbf{2049} & 4098 & 6147 & 8822 \\ 
\textbf{15} & & \textbf{1366} & 2732 & 4098 & 5881 \\ 
\textbf{20} & & \textbf{1025} & \textbf{2049} & 3074 & 4411 \\
\textbf{24} & & \textbf{854} & \textbf{1708} & 2561 & 3676 \\ 
30 & & 683 & 1366 & 2049 & 2941 \\ 
\hline
\end{tabular}
\end{table}

The study of \cite{Kargaltsev2017} provides a list of kick velocities for bow-shock \ac{PWN}e (see their Table\,1), which varies from 60\,km\,s$^{-1}$ to around 2,000\,km\,s$^{-1}$. The upper limit of this range for the projected velocity of Potoroo suggests that a minimum age of 10,000\,years is consistent with previous estimates. 

We also investigated the pulsar velocity for the case of Potoroo association with \ac{SNR}~G338.1+0.4 \citep{1970AuJPA..14..133S,1996A&AS..118..329W}, since the trail emission appears to point towards the \ac{SNR} geometrical centre (Figure\,\ref{fig:rgb}). Distances to G338.1+0.4 were reported as 6.4\,kpc \citep{1970AuJPA..14..133S}, 9.9\,kpc \citep{1998ApJ...504..761C}, 11.9\,kpc \citep{2007Ap&SS.307..423S} and 6.2\,kpc \citep{2014SerAJ.189...25P}, all derived using the brightness--to--diameter ($\Sigma$--D) relation. Given the large angular separation of approximately 50\arcmin, it is highly unlikely for G338.1+0.4 to be the parent \ac{SNR} of Potoroo even if they were as close as 6.2\,kpc (Table\,\ref{tab:speed}). 



Our analysis suggests that the progenitor star likely exploded within a distance no greater than twice the size of the Potoroo tail, with the natal kick velocity well exceeding 1000\,km\,s$^{-1}$. Identifying a parent candidate in the chosen area is challenging due to the extreme density of \HII~regions. The high concentration of thermal emission in these regions may hinder the detection of any emission from the \ac{SNR}, as discussed in Section\,\ref{subsec:snr}. Therefore, identifying the origin of Potoroo will require further observations and analyses.

\subsection{Potoroo's Tail Lengths}
\label{subsec:tail}

The observed sizes of Potoroo in radio and X-ray data are approximately 7.2\arcmin\ and 0.67\arcmin, respectively. The size difference is attributed to the synchrotron ageing of relativistic particles generated by a fast-moving pulsar. The X-ray-emitting particles have a shorter lifespan and smaller spatial extent compared to the radio-emitting particles. In contrast, the radio-emitting particles persist for a longer time and form a larger, more diffuse structure that extends further away from the X-ray emission region. 

To estimate the physical length of Potoroo's tails, we converted the angular sizes to minimal lengths of 21\,pc and 2\,pc for radio and X-ray sources, respectively. However, we believe that the tails may be longer than these estimates. The X-ray emission is only detectable above 2\,keV likely due to the severe absorbing column and the emission from a possibly longer tail is blended into the background. On the other hand, the radio tail is well-defined until it reaches the complexity of the \ac{ISM}, beyond which we cannot confidently identify the Potoroo tail.

Among all the observed pulsar tails, Potoroo stands out with a projected radio length of 21\,pc, making it the longest observed \ac{PWN}. The \ac{PWN} of PSR\,J1437$-$5959, associated with the Frying Pan \ac{SNR}, has a similar radio extent of over 20\,pc \citep{2009ApJ...703L..55C}, but is not visible in X-rays. The projected velocity of the driving pulsar is $\sim$1000\,km\,s$^{-1}$ \citep{2012ApJ...746..105N}. The Mouse \ac{PWN} is the next in size and has radio and X-ray tails spanning 17 and 1\,pc, respectively \citep{2004ApJ...616..383G}, and a pulsar velocity of 300\,km\,s$^{-1}$ \citep{2009ApJ...706.1316H}. The G319.9$-$0.7 \ac{PWN} has the longest X-ray tail of 7\,pc, which is also relatively big compared to its 10\,pc radio tail \citep{2010ApJ...712..596N}. The velocity of the central pulsar J1509$-$5850 is $\sim$300\,km\,s$^{-1}$. As noted earlier, the appearance of \ac{PWN} is strongly influenced by the physical properties of their surrounding environment, such as density and magnetic field. Hence, a weak correlation between the pulsar velocity and the radio length is not unexpected.

\subsection{Radio and X-Ray Spectral Analysis}
\label{subsec:break}

The spectral map of Potoroo shows a flat radio continuum spectrum near the pulsar due to younger and more energetic electrons, while the spectrum becomes steeper as the distance from the pulsar increases, where synchrotron radiation originates from the older electron population. To examine the spectral variations of Potoroo, we obtained spectral indices from three different regions: the main body, the diffuse part of the tail, and the entire source area (see Figure\,\ref{fig:spectral_index} andTable\,\ref{tab:index}). 

While the main body of Potoroo exhibits a typical flat spectral index of $-$0.21, its overall spectral index is much steeper than that of any other observed \ac{PWN}. The dominance of its diffuse tail is evident, but its remarkable length coupled with the unusually steep spectrum raises intriguing questions. Taken together, these observed properties provide additional support for classifying Potoroo as a mature \ac{PWN}. 

When a \ac{PWN} interacts with the reverse shock of its parent \ac{SNR}, the relativistic particles within the nebula undergo re-acceleration. This process significantly enhances the brightness of the nebula, particularly at radio frequencies. Simultaneously, the magnetic field of the \ac{PWN} gets compressed by the shock. As synchrotron radiation is emitted when charged particles spiral in a magnetic field, the frequency of this radiation depends on both the energy of particles and the strength of the magnetic field. A stronger magnetic field results in higher frequency radiation leading to a rapid synchrotron burn-off and steepening of the spectral index. The steepest spectral indices known are associated with \ac{PWN}e that have collided with the reverse shock. However, even in these cases, the spectral indices do not exceed $-$0.7, e.g. $\alpha=-0.69$ \cite[G141.2+5+0,][]{2014ApJ...784L..26K} and $\alpha=-0.62$ \cite[G76.9+1.0,][]{1993A&A...276..522L}. 

\begin{table}[hbt!]
\caption{ Spectral $\alpha$ and photon $\Gamma$ indices of Potoroo. The spectral indices are calculated from the spectral map generated using data at 944, 1284 and 1368\,MHz, as described in Section\,\ref{subsec:spectral_analysis}. The photon index is derived by fitting combined \chandra\ and \textit{XMM-Newton} data within the energy range of $2-10$\,keV.}
\label{tab:index}
\begin{tabular}{l c c c c }
\toprule
\makecell{} & \makecell{full source} & \makecell{main body} & \makecell{diffuse tail} & \makecell{}\\
\hline
\hline
$\alpha$ & $-$1.26$\pm$0.04 & $-$0.21$\pm$0.02 & $-$1.42$\pm$0.04 &  \\  
$\Gamma_{\textit{XMM}+\chandra}$ & 1.1$\pm$0.7\footnote{Reference: \cite{2014ApJ...787..129J}.} & $-$ & $-$ &  \\ 
\end{tabular}
\end{table}

In the \chandra\ observations, \cite{2014ApJ...787..129J} found a similar trend of the spectrum softening with increasing distance from the Potoroo pulsar (see their Figure 2). The authors divided the X-ray source emission into three narrow bands with equal counts. By comparing the softest image with the hardest, they observed that the emission tends to clump towards the point-like source for higher energies, thereby enhancing the prominence of the point-like source and reducing the tail emission.

At X-ray energies and beyond, synchrotron emission is commonly modeled as a power-law distribution of photons: N$_{E}\propto E^{-\Gamma}$, where N$_{E}$ is the number of photons with energies E and $\Gamma=1-\alpha$ is the photon index. In the context of \ac{PWN}, the typical value for $\Gamma$ is around 2 \citep{2006ARA&A..44...17G}, and the index tends to increase (softens, steepens) with greater distances from the pulsar.

The photon index of Potoroo of 1.1 is taken from the work of \cite{2014ApJ...787..129J}, and also listed in Table\,\ref{tab:index}. The authors used data from \textit{XMM-Newton} to supplement the limited photon statistics of the \chandra\ data and refine constraints on the spectral parameters. Due to the limited angular resolution of \textit{XMM-Newton}, the point source and extended emission of Potoroo could not be resolved, so only the full source spectrum was fitted. The calculated photon index of 1.1 is much harder than typical X-ray \ac{PWN} spectra. This hardness is expected, given that the source is detectable only above 2\,keV and lacks any softening towards lower energies.

The collision between \ac{PWN} and the reverse shock leads to the efficient blending of thermal and non-thermal gases \citep{2001ApJ...563..806B}, which is expected to produce thermal soft X-ray emissions. \cite{2014ApJ...787..129J} attempt to fit the Potoroo spectra with a thermal model resulted in encountering unconstrained parameters and unacceptable fit statistics.

 The complete understanding of the Potoroo’s spectrum remains open. Further investigations using higher frequency radio bands, deeper X-ray data and potential $\gamma$ detections are necessary to determine its unique characteristics. However, due to the high absorption in the line of sight toward the Norma spiral arm, it is unlikely that searching for thermal X-ray emission or a bow-shock in H$\alpha$ will yield any significant results.

\subsection{Properties of PSR~J1638--4713}
\label{subsec:pulsar_discussion}

It is well known that \acp{NS} move at higher velocities than their progenitor stars~\citep{1970ApJ...160..979G}. In fact, a significant number of isolated \acp{NS} have velocities exceeding 1000\,km\,s$^{-1}$ \citep[e.g.,][]{2005MNRAS.360..974H,2019ApJ...875..100D}. These high velocities are believed to result from natal kicks that occur during the \acp{NS}' formation. The natal kicks could be caused by various mechanisms, including asymmetric explosions triggered by hydrodynamical perturbations in the supernova core~\citep[e.g.,][]{2000ApJ...535..402L}, or asymmetric neutrino emission in the presence of super-strong magnetic fields ($B>10^{15}$\,G) in the proto-neutron star~\citep[e.g.,][and references therein]{1999ApJ...519..745A,1999PhRvD..60d3001A}. Therefore, the discovery of fast-moving pulsars might shed 
light on the physics of supernovae. 
As discussed in Section~\ref{subsec:distance}, the transverse velocity of \psr\ is likely to be higher than 1000\,km\,s$^{-1}$, which makes it among the fastest moving \acp{NS}. The relatively strong radio continuum emission from the \ac{PWN} also opens up the opportunity to place constraints on the proper motion and parallax of \psr~\citep{2019ApJ...875..100D}. These measurements will be crucial for us to understand the origin and properties of Potoroo.

\begin{figure*}[hbt!]
\centering
\includegraphics[width=0.65\textwidth]{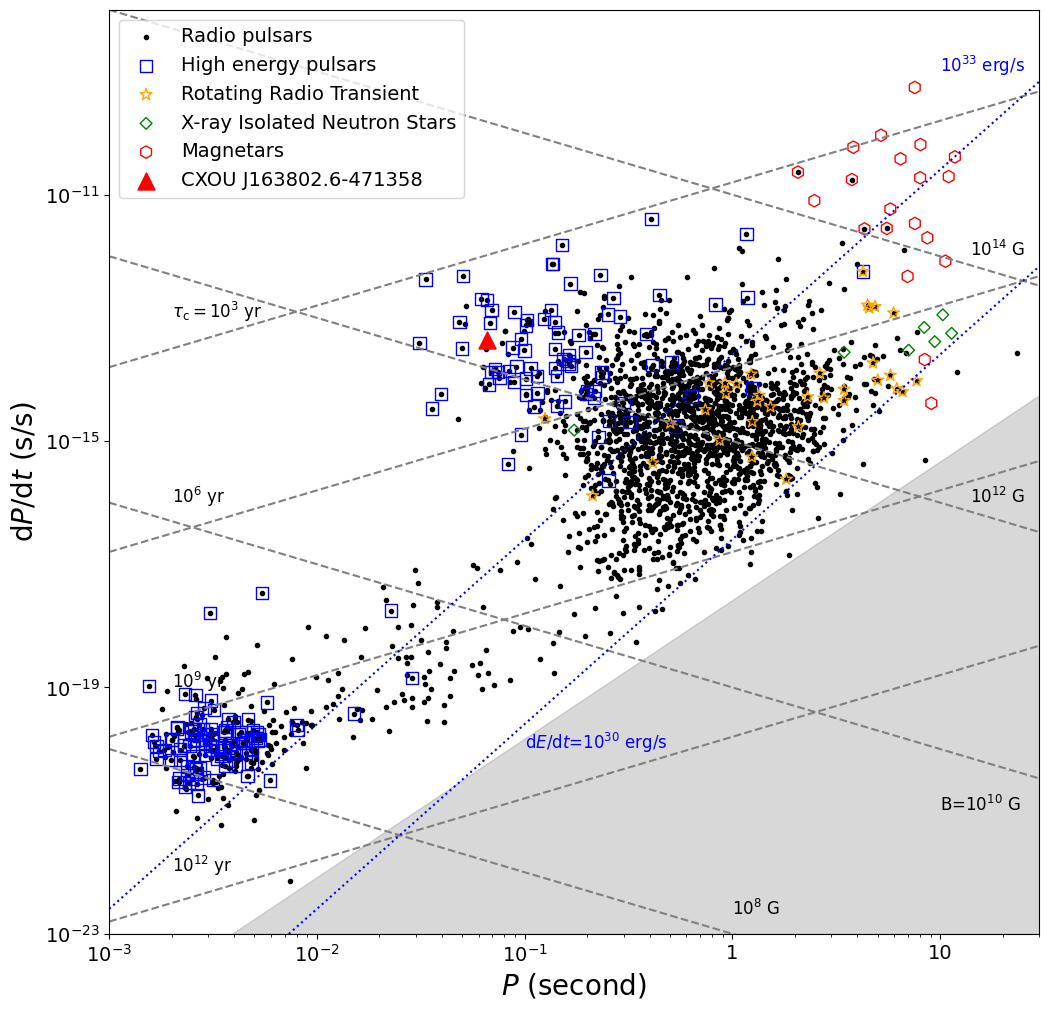}
\caption{The spin period ($P$) versus the time derivative of the spin period ($\dot{P}$) for known pulsars (the so-called `P-Pdot' diagram). Different types of pulsars are shown with different markers and \psr\ (CXOU~J163802.6$-$471358) is shown as a red triangle. We plot contours of characteristic ages, magnetic fields and spin-down luminosity based on the canonical pulsar spin-down model~\citep{2004hpa..book.....L}. }
\label{fig:ppdot}
\end{figure*}

The spin parameters of \psr\ are consistent with those of young, energetic pulsars. In Figure\,\ref{fig:ppdot}, we show the spin period ($P$) and the time derivative of the spin period ($\dot{P}$) for all known pulsars listed in the ATNF Pulsar Catalogue~\citep{2005AJ....129.1993M}. \psr\ is shown as a red triangle and it is located in a region where young pulsars with high energy emissions are. Based on our current estimate of the $\dot{P}$, the spin-down luminosity of \psr\ is estimated to be $\sim6.1\times10^{36}$\,erg\,s$^{-1}$, revealing an energy loss rate that is among the highest for supersonic pulsars associated with \ac{PWN}~\cite[Table 1,][]{Kargaltsev2017}. We are currently regularly observing \psr\ with the Parkes telescope, and a longer pulsar timing baseline will enable us to precisely measure the spin and astrometric parameters of this pulsar. Better timing models of this pulsar will also enable us to search for high-energy pulsation (e.g., X-ray, $\gamma$-ray) in the future.

\section{Conclusion}
\label{sec:conclusion}

We report the discovery of a bow-shock \ac{PWN}, named Potoroo, in the new radio continuum surveys obtained by \ac{ASKAP} and MeerKAT, and the detection of the associated \psr\ using Parkes observations. We also compared our results with the previous study of the X-ray source detected during the Norma spiral arm \chandra\ survey. The object exhibits distinctive cometary morphology in both radio and X-ray domains, suggesting pulsar leading the \ac{PWN} and travelling supersonically through the ambient medium. We estimated a large projected velocity of \psr, well exceeding 1000\,km\,s$^{-1}$. The pulsar was identified above 3\,GHz at a very high \ac{DM} of 1553\,\dmunits. Potoroo's radio size is longer than its X-ray size by a factor of 10. With a distance of at least 10\,kpc, the physical sizes are approximately 21\,pc and 2\,pc for radio and X-ray sources, respectively. This makes Potoroo one of the longest observed radio \ac{PWN} trail to date.  

Our analysis of Potoroo has revealed an unusually steep overall radio spectrum ($\alpha=-1.27$), which falls far below the typical values for \ac{PWN}e. However, focusing only on the high signal-to-noise ratio part of the nebula, we found that its $\alpha$ value is consistent with the majority of \ac{PWN}e. We attribute the steep overall spectral index to the interaction of the parent \ac{SN} reverse shock with the \ac{PWN}, which results in an increase of the magnetic field of the \ac{PWN} and severe synchrotron burn-off of the highest energy electrons. This interaction also makes the radio nebula brighter, and along with massive absorption of the surrounding medium, explains the significant length difference observed between the radio and X-ray sources.

Multi-frequency studies of Potoroo are essential to further understand the physics of this remarkable object and its evolution in a very complex environment. In particular, our polarisation analysis revealed highly ordered polarisation vectors that run parallel to the Potoroo's tail orientation, but additional high-frequency polarimetry is required to constrain the magnetic field. Dedicated Chandra longer exposure observations are necessary in order to investigate the nature of the misaligned outflow and improve our image of the flow structure of the X-ray particles. The parent \ac{SNR} is not known so far, and to deeper investigate the origin of Potoroo, we intend to study the distribution of the interstellar medium in the surrounding environment through molecular lines. Based on the \psr\ spin-down and characteristic age, we anticipate detecting gamma-ray photons from Potoroo. If TeV emission is detected, we can accurately measure the magnetic field's strength by jointly modelling the synchrotron and inverse Compton scattering emission.

\section*{Acknowledgement}

This scientific work uses data obtained from Inyarrimanha Ilgari Bundara, the CSIRO Murchison Radio-astronomy Observatory. We acknowledge the Wajarri Yamaji People as the Traditional Owners and native title holders of the Observatory site. CSIRO’s \ac{ASKAP} radio telescope is part of the \ac{ATNF}\footnote{\url{https://ror.org/05qajvd42}}. Operation of \ac{ASKAP} is funded by the Australian Government with support from the National Collaborative Research Infrastructure Strategy. \ac{ASKAP} uses the resources of the Pawsey Supercomputing Research Centre. Establishment of ASKAP, Inyarrimanha Ilgari Bundara, the CSIRO Murchison Radio-astronomy Observatory and the Pawsey Supercomputing Research Centre are initiatives of the Australian Government, with support from the Government of Western Australia and the Science and Industry Endowment Fund. The Parkes radio telescope (recently given the Indigenous Wiradjuri name Murriyang) is also part of \ac{ATNF} which is funded by the Australian Government for operation as a National Facility managed by CSIRO. We acknowledge the Wiradjuri people as the Traditional Owners of this observatory site. 
 
The MeerKAT telescope is operated by the South African Radio Astronomy Observatory, which is a facility of the National Research Foundation, an agency of the South Africa Department of Science and Innovation. 

The National Radio Astronomy Observatory is a facility of the US National Science Foundation, operated under a cooperative agreement by Associated Universities, Inc.
 
This publication also makes use of data products from the \ac{CDA} and software provided by the Chandra X-ray Center (CXC), as well as from the Wide-field Infrared Survey Explorer, a joint project of the University of California, Los Angeles, and the Jet Propulsion Laboratory/California Institute of Technology, that is funded by the \ac{NASA}.  The authors acknowledge the use of the \ac{NASA}’s Astrophysics Data System Bibliographic Services\footnote{\url{http://adsabs.harvard.edu}} and Simbad Astronomical Database\footnote{\url{http://simbad.u-strasbg.fr/simbad}}, operated at the CDS, Strasbourg, France. Carta\footnote{\url{https://cartavis.org}} and SAOImage DS9\footnote{\url{https://sites.google.com/cfa.harvard.edu/saoimageds9}} \citep{ds9} tools are extensively used for image display and visualisation.

We thank Fernando Camilo for the early-stage paper discussion and for providing the MeerKAT data. We also thank Naomi van Jaarsveld, Benjamin Stappers and Vanessa McBridge for conducting the initial Parkes observation of Potoroo and generously sharing their notes on data reduction.

S.L., M.D.F., and G.R. acknowledge the \ac{ARC} funding through grant DP200100784. S.L. also acknowledges support from the Ministry of Education, Science and Technological Development of the Republic of Serbia through contract number 451-03-47/2023-01/200002. S.D. is the recipient of an \ac{ARC} Discovery Early Career Award (DE210101738) funded by the Australian Government.

 We thank an anonymous referee for comments and suggestions that greatly improved our paper.

\section*{Data Availability}

This study made use of archival \ac{ASKAP} data obtained from the \ac{CASDA}\footnote{\url{https://research.csiro.au/casda}}. The observations from the Parkes radio telescope are accessible through the same portal but after an 18-month embargo period. The corresponding author can provide the MeerKAT data upon reasonable request. The X-ray imagining results are based on data from the \ac{CDA}\footnote{\url{https://cxc.harvard.edu/cda}}. \ac{WISE} maps were obtained via the SkyView Virtual Observatory\footnote{\url{https://skyview.gsfc.nasa.gov/current/cgi/titlepage.pl}} server, which is supported by \ac{NASA}'s High Energy Astrophysics Science Archive Research Center (HEASARC). This study also drew information from the \ac{WISE} Catalog of Galactic \HII~Regions\footnote{\url{http://astro.phys.wvu.edu/wise}} and the \ac{ATNF} Pulsar Catalogue\footnote{\url{https://www.atnf.csiro.au/research/pulsar/psrcat}}.
